\documentclass[twocolumn]{aastex701}

\usepackage{amsmath}
\usepackage{mathtools}

\begin{document}

\title{Nova-driven Dynamical Friction Regulates the Survival of Cataclysmic Variables}
\shorttitle{Nova-drag CAML in CVs}
\shortauthors{Dodon et al.}

\author[orcid=0009-0003-7483-0090,gname=Vladislav,sname=Dodon]{Vladislav Dodon}
\affiliation{School of Astronomy and Space Science, Nanjing University, Nanjing, 210023, People's Republic of China}
\affiliation{Kazan Federal University, Kremlevskaya Str.18, 420008, Kazan, Russia}
\email[show]{vidodon@kpfu.ru}  

\author[orcid=0000-0002-0584-8145,gname=Xiang-Dong, sname=Li]{Xiang-Dong Li} 
\affiliation{School of Astronomy and Space Science, Nanjing University, Nanjing, 210023, People's Republic of China}
\affiliation{Key Laboratory of Modern Astronomy and Astrophysics, Ministry of Education (Nanjing University), Nanjing, 210023, People's Republic of China}
\affiliation{Yingtian Laboratory of Space Astronomy Science and Engineering, Nanjing 210033, P. R. China}
\email{lixd@nju.edu.cn}

\author[orcid=0000-0003-1535-0866,gname=Diogo, sname=Belloni]{Diogo Belloni} 
\affiliation{International Centre of Supernovae (ICESUN), Yunnan Key Laboratory of Supernova Research, Yunnan Observatories, Chinese Academy of Sciences (CAS), Kunming, 650216, People's Republic of China}
\email{diogo@ynao.ac.cn}

\author[orcid=0000-0003-3903-8009,gname=Matthias, sname=Schreiber]{Matthias R. Schreiber} 
\affiliation{Departamento de F{\'i}sica, Universidad T{\'e}cnica Federico Santa Mar{\'i}a, Av. Espa{\~n}a 1680, Valpara{\'i}so, Chile}
\email{matthias.schreiber@usm.cl}

\author[orcid=0000-0001-5778-2355,gname=Ilkham, sname=Galiullin]{Ilkham Galiullin} 
\affiliation{Kazan Federal University, Kremlevskaya Str.18, 420008, Kazan, Russia}
\email{ilhigaliullin@kpfu.ru}

\author[orcid=0000-0002-7909-4171,gname=Zhengwei, sname=Liu]{Zhengwei Liu} 
\affiliation{International Centre of Supernovae (ICESUN), Yunnan Key Laboratory of Supernova Research, Yunnan Observatories, Chinese Academy of Sciences (CAS), Kunming, 650216, People's Republic of China}
\email{zwliu@ynao.ac.cn}

\begin{abstract}
The evolution and ultimate fate of interacting binaries are governed by mass exchange and orbital angular momentum loss, yet the physical origin of some of the angular momentum loss required by observations remains elusive. Cataclysmic variables (CVs) provide a particularly striking example: their white dwarfs (WDs) are systematically more massive than those in detached progenitors, implying that binaries containing low-mass WDs are preferentially removed from the CV population. Population models can reproduce this effect by invoking empirical angular momentum loss prescriptions whose strength increases towards lower WD masses, triggering mergers due to dynamically unstable mass transfer. However, the physical origin of this mass dependence has remained unexplained for more than a decade. Here we show that dynamical friction exerted by nova ejecta on the donor star naturally provides the missing mechanism. We derive the angular momentum loss during nova eruptions from gaseous dynamical friction and evaluate it using ejecta-velocity prescriptions motivated by nova evolution calculations. Because nova ejecta are substantially slower for low-mass WDs, they are more strongly gravitationally focused by the companion, producing enhanced drag and preferentially driving these systems towards unstable mass transfer. Our binary population synthesis shows that this physically motivated nova-drag prescription reproduces remarkably well the observed CV WD mass distribution. Our results reveal how short-lived nova eruptions regulate the secular evolution and survival of compact binaries, providing a physical link between transient mass-loss events and the long-term demographics of interacting binary populations.
\end{abstract}


\section{Introduction} 

Cataclysmic variables (CVs) are compact interacting binaries in which a white dwarf (WD) accretes matter from a Roche-lobe-filling low-mass donor star \citep{2003cvs..book.....W}. CVs have been studied for decades as natural laboratories for accretion physics and interacting-binary evolution \citep{2026SSRv..222...32S}. Yet despite this long history, our understanding of CV evolution remains in a state of crisis \citep[see][for a review]{2020AdSpR..66.1080Z}. 

A growing body of observational evidence has converged on a triple discrepancy that standard evolutionary models \citep[e.g.][]{1993A&A...271..149K, 2001ApJ...550..897H, 2011ApJS..194...28K} cannot explain: the WD mass, orbital-period distribution, and space density problems \citep{2009MNRAS.397.2170G, 2011A&A...536A..42Z, 2012MNRAS.419.1442P, 2020MNRAS.494.3799P}. These three anomalies are not independent. They share a common root because standard theory predicts a large population of short-period CVs containing low-mass WDs that observations simply do not find. 

The most successful resolution proposed to date is the empirical consequential angular momentum loss model (eCAML; \citealt{2016MNRAS.455L..16S}). It produces enhanced angular momentum loss (AML) and destabilizes mass transfer in systems containing low-mass WDs, causing many of them to merge. This preferentially removes low-mass WDs from the surviving CV population, bringing the predicted WD mass distribution into agreement with observations and also alleviating the discrepancies in the orbital-period distribution and space density of CVs \citep{2018MNRAS.478.5626B, 2022MNRAS.510.6110P}. 

Despite these successes, eCAML remains an empirical prescription without an established physical origin. \citet{2016MNRAS.455L..16S} suggested that the enhanced AML might result during classical nova eruptions from friction between the expanding nova ejecta and the binary, particularly in systems containing low-mass WDs, which are expected to produce slower ejecta \citep{1995ApJ...445..789P,2005ApJ...623..398Y}. Frictional angular momentum loss (FAML) during nova eruptions has been investigated in several studies \citep[e.g.,][]{1986ApJ...305..251M, 1986ApJ...311..163S, 1991A&A...246...84L, 1998MNRAS.297..633S}. Most notably, \citet{2019ApJ...870...22L} implemented the FAML prescription of \citet{1998MNRAS.297..633S} in detailed evolutionary calculations and found that FAML is generally too weak to influence CV evolution. Previous FAML prescriptions approximated the frictional interaction using a ram-pressure drag force with the interaction scale set by the donor's radius. 

However, the donor's gravity can deflect ejecta passing well beyond its surface, producing a trailing overdense wake whose gravitational pull exerts an additional drag force on the donor. For sufficiently slow ejecta, the gravitational focusing radius can substantially exceed the donor radius, allowing much stronger friction. This physical picture is motivated by the gaseous dynamical friction formalism of \citet{1999ApJ...513..252O}, which extends the classical collisionless result of \citet{1943ApJ....97..255C} to a point mass perturber moving rectilinearly through a gaseous medium. Subsequent studies extended its application to circular trajectories and finite-size perturbers \citep{2007ApJ...665..432K, 2013ApJ...775...72B}.

Here we present a new prescription for consequential angular momentum loss that accounts for nova-driven dynamical friction, which we call the nova-drag model. We combine this prescription with ejecta velocities motivated by nova calculations and use binary population synthesis to evaluate its effects on CV evolution. The model predicts surviving CV populations similar to those obtained with eCAML and reproduces the observed WD mass distribution.

\section{Consequential angular momentum loss in CVs}
\label{sec:CAML}

Classical nova eruptions, powered by thermonuclear runaways in the accreted hydrogen-rich envelope on the WD surface, recur throughout CV evolution \citep{2013ApJ...777..136W, 2021ARA&A..59..391C}. The ejected material carries away orbital angular momentum, producing an additional AML channel known as consequential angular momentum loss (CAML; \citealt{1995ApJ...439..330K}).

CAML is commonly parametrized as
\begin{equation}
\label{eq:Jdot_CAML}
    \frac{\dot J_\mathrm{CAML}}{J_\mathrm{orb}} = \nu \frac{\dot M_2}{M_2},
\end{equation}
\noindent where $J_\mathrm{orb}$ is the orbital angular momentum, $M_2$ is the donor mass, and $\nu$ is a dimensionless coefficient describing how much specific angular momentum is carried away by the ejected material. We assume that mass transfer in CVs is fully non-conservative, so that all material accreted by the WD is subsequently expelled during nova eruptions. This approximation is broadly consistent with nova calculations \citep{2005ApJ...623..398Y, 2005A&A...439.1061S}.

One contribution to CAML arises from mass loss through nova ejecta. We assume that, before interacting with the donor, ejecta launched from the vicinity of the WD carry the WD's specific orbital angular momentum. This is commonly referred to as classical CAML, with the corresponding coefficient given by \citep{1995ApJ...439..330K}
\begin{equation}
\label{eq:nu_iso}
\nu_\mathrm{iso}=\frac{q^2}{1+q},
\end{equation}
where $q=M_2/M_\mathrm{WD}$. This prescription is equivalent to the isotropic re-emission model widely used in compact binary evolution \citep{1997A&A...327..620S}.

Additional AML comes from friction between the nova ejecta and the donor. This interaction transfers orbital angular momentum to the ejecta in addition to that carried at launch. The total CAML coefficient is therefore
\begin{equation}
\label{eq:nu_tot}
\nu_\mathrm{tot}=\nu_\mathrm{iso}+\nu_\mathrm{drag},
\end{equation}
where $\nu_\mathrm{drag}$ describes the additional loss due to dynamical friction.

Throughout the paper, we compare our results with the eCAML prescription \citep{2016MNRAS.455L..16S}:
\begin{equation}
\label{eq:nu_eCAML}
\nu_\mathrm{eCAML}=\frac{C}{M_\mathrm{WD}},
\end{equation}
where $C\sim0.3\,M_\odot$. This prescription represents the total CAML coefficient, replacing $\nu_\mathrm{tot}$ rather than adding to $\nu_\mathrm{iso}$.

\section{The nova-drag model}
\label{sec:model}

\begin{figure}
    \centering
    \includegraphics[width=0.99\linewidth]{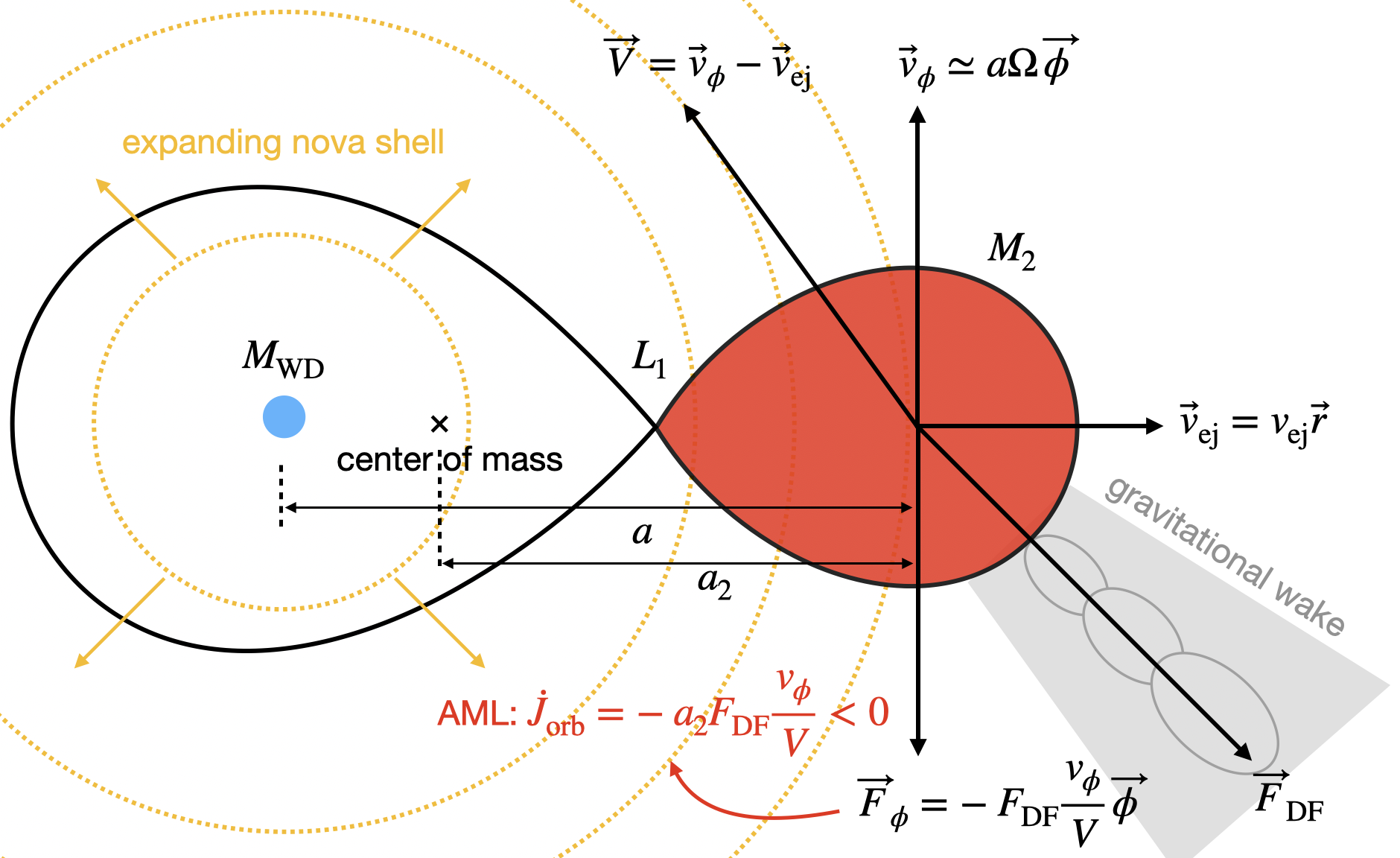}
    \caption{Schematic geometry of the dynamical friction interaction between a radially expanding nova shell and the Roche-lobe-filling donor, showing the kinematics and drag force responsible for orbital angular momentum loss.}
    \label{fig:derive}
\end{figure}

We first consider the angular momentum extracted during a single nova eruption. To obtain an analytic estimate, we idealize the ejecta crossing the binary as a spherical shell with density $\rho(t)$ and radial velocity $v_\mathrm{ej}$, given by
\begin{equation}
\label{eq:v_ej}
v_\mathrm{ej}=v_0\left(\frac{M_\mathrm{WD}}{0.8\,M_\odot}\right)^p,
\end{equation}
\noindent where $p=3$, indicating a relatively steep dependence of the ejecta velocity on the WD mass, and $v_0=400\,\mathrm{km\,s^{-1}}$, based on the nova calculations of \citet{2005ApJ...623..398Y}. This relation provides a representative approximation to the expansion velocities averaged over the nova mass-loss phase (see Appendix~\ref{app:velocity} for more details).

We show in Figure~\ref{fig:derive} the geometry and conventions adopted in the derivation (see also \citealt{1998MNRAS.297..633S} for a similar derivation of FAML). For a circular binary with separation $a$, the orbital angular momentum is $J_\mathrm{orb} = M_\mathrm{WD}M_2a^2\Omega / (M_\mathrm{WD}+M_2)$ and the orbital frequency is $\Omega = \sqrt{G(M_\mathrm{WD}+M_2) / a^3}$. The donor moves through the ejecta and gravitationally focuses the surrounding gas, producing a trailing overdense wake. The gravitational pull of this wake exerts a drag force on the donor and removes orbital angular momentum from the binary. 

The magnitude of the drag force on a perturber of mass $M_2$ moving through gas of density $\rho$ with relative velocity $V$ is \citep{1999ApJ...513..252O}
\begin{equation}
\label{eq:O99_force}
F_\mathrm{DF} = 4\pi I(\mathcal{M}) \frac{G^2M_2^2\rho}{V^2},
\end{equation}
where $I(\mathcal{M})$ is a scale factor which depends on the Mach number ($\mathcal{M}=V/c_\mathrm{s}$). 

The force acts opposite to the relative velocity between the donor and the ejecta, which can be expressed as
\begin{equation}
\label{eq:relative_velocity}
V \simeq \sqrt{v_\mathrm{ej}^2+v_\phi^2},
\end{equation}
where $v_\phi$ is the tangential component of the donor--ejecta relative velocity. For ejecta launched from the vicinity of the WD, we approximate this component by the relative orbital velocity of the two stars, $v_\phi\simeq a\Omega$. Only the tangential component of the drag force participates in extracting orbital angular momentum:
\begin{equation}
\label{eq:tangent_force}
F_\phi = F_\mathrm{DF}\frac{v_\phi}{V}.
\end{equation}

The drag acts only while the nova shell crosses the binary. The angular-momentum impulse during one eruption is therefore
\begin{equation}
\label{eq:deltaJ_drag}
|\Delta J_\mathrm{drag}| = \int a_2 F_\phi(t)\,dt,
\end{equation}
where $a_2=M_\mathrm{WD}a/(M_\mathrm{WD}+M_2)$ is the distance of the donor from the centre of mass. For an ejecta mass $\Delta M_\mathrm{ej}$ crossing the donor orbit, the shell column satisfies
\begin{equation}
\label{eq:shell_column}
\int \rho(t)\,dt = \frac{\Delta M_\mathrm{ej}}{4\pi a^2v_\mathrm{ej}}.
\end{equation}
Substituting into Eq.~\ref{eq:deltaJ_drag}, and assuming that $a_2$ and the velocities do not vary strongly while the shell crosses the binary, gives
\begin{equation}
\label{eq:deltaJ_result}
|\Delta J_\mathrm{drag}| = a_2 I(\mathcal{M}) \frac{G^2M_2^2v_\phi}{a^2v_\mathrm{ej}V^3} \Delta M_\mathrm{ej}.
\end{equation}
This gives the AML for a single nova event, which can be used directly in detailed binary evolutionary calculations to test the discontinuous effect that nova mass loss has on CV evolution (see for example \citealt{2024ApJ...977...34T}).

To connect the impulsive AML given in Eq.~\ref{eq:deltaJ_result} with the secular CAML parametrization in Eq.~\ref{eq:Jdot_CAML}, we average over many nova cycles. Although the drag acts only while the nova shell crosses the binary, the cumulative effect of repeated, short-lived drag impulses can be written in the standard CAML form by defining an effective drag coefficient:\footnote{This assumes that the WD ejects all the accreted mass during nova eruptions, so that $\Delta M_\mathrm{ej}\simeq-\Delta M_2$. For a recurrence time $t_\mathrm{rec}$, the cycle-averaged torque is then $\langle \dot J_\mathrm{drag} \rangle = \Delta J_\mathrm{drag}/t_\mathrm{rec}$, while the corresponding secular mass-transfer rate satisfies $|\dot M_2| \simeq \Delta M_\mathrm{ej}/t_\mathrm{rec}$. Thus, the recurrence time cancels in the definition of the secular CAML coefficient. The use of a continuous CAML prescription is also supported by the analytical and numerical calculations of \citet{1998MNRAS.297..633S}, who found that the averaged treatment reproduces the mass-transfer stability criterion of resolved nova cycles.}
\begin{equation}
\label{eq:nu_drag_def}
\nu_\mathrm{drag}
=
\frac{M_2}{J_\mathrm{orb}}
\frac{|\Delta J_\mathrm{drag}|}{\Delta M_\mathrm{ej}} .
\end{equation}

Substituting Eq.~\ref{eq:deltaJ_result} into Eq.~\ref{eq:nu_drag_def} gives
\begin{equation}
\label{eq:nu_drag_full}
\nu_\mathrm{drag} = I(\mathcal{M}) \frac{G^2M_2^2}{a^2v_\mathrm{ej}V^3}.
\end{equation}
The dependence is set by the gravitational focusing strength of the donor and by the kinematics of the ejecta. Introducing the gravitational focusing radius as $R_\mathrm{g}\approx 2GM_2/V^2$ appropriate for supersonic motion $c_\mathrm{s}\ll V$, the drag-CAML coefficient becomes
\begin{equation}
\label{eq:nu_drag_final}
\nu_\mathrm{drag} = \frac{I(\mathcal{M})}{4}
\left(\frac{R_\mathrm{g}}{a}\right)^2
\frac{V}{v_\mathrm{ej}}.
\end{equation}

The remaining ingredient in Eq.~\ref{eq:nu_drag_final} is the scale factor $I(\mathcal{M})$. \citet{1999ApJ...513..252O} obtained
\begin{equation}
\label{eq:scale_factor_I}
I(\mathcal{M})
=
\begin{cases}
\dfrac{1}{2}\ln\!\left(\dfrac{1+\mathcal{M}}{1-\mathcal{M}}\right)-\mathcal{M},
& \mathcal{M}<1, \\[1.2ex]
\ln\!\left(\dfrac{b_{\max}}{r_{\min}}\right)
+\dfrac{1}{2}\ln\!\left(1-\mathcal{M}^{-2}\right),
& \mathcal{M}>1 .
\end{cases}
\end{equation}
Here $b_{\max}$ is the largest scale over which the surrounding gaseous medium participates in the interaction, and $r_{\min}$ is the effective perturber size. In the nova-ejecta case, the motion is expected to be supersonic. For example, over a wide temperature range, $T_\mathrm{ej}\sim10^4$--$10^6\,\mathrm{K}$, the ionized gas will have $c_\mathrm{s}\sim15\text{--}150\,\mathrm{km\,s^{-1}}$. The orbital velocity in CVs is typically several hundred to a thousand $\mathrm{km\,s^{-1}}$. Thus $\mathcal{M}=V/c_\mathrm{s}>1$ for the full parameter range, and usually $\mathcal{M}\gg1$. In this limit, the second term in the supersonic expression is small, so that $I(\mathcal{M})\simeq\ln(b_{\max}/r_{\min})$.

The original calculation of \citet{1999ApJ...513..252O} within linear perturbation theory assumed a point mass moving on a straight trajectory. Later, \citet{2007ApJ...665..432K} showed that the same prescription remains a good approximation for circular orbits if the outer cut-off is taken to be $b_{\max}\simeq2a$. Then, \citet{2013ApJ...775...72B} showed that for an extended object the same formulas remain viable if we take $r_\mathrm{min} = \mathrm{max}(2.25R_2,R_\mathrm{g})$ (consistent as well with \citealt{1999ApJ...522L..35S}). 

Although these prescriptions have been validated for gaseous dynamical friction in more general configurations, their application to the rapidly expanding and potentially inhomogeneous nova ejecta remains approximate. The limitations of our analytical approach are discussed in Appendix~\ref{app:assume}. We consider the derived prescription a reasonable first-order approximation for studying secular CV evolution. It may also find applications to other classes of compact binaries in which ejected material interacts with a companion \citep[e.g.,][]{2015ApJ...805L...6S}.

\begin{figure*}
    \centering
    \includegraphics[width=0.99\linewidth]{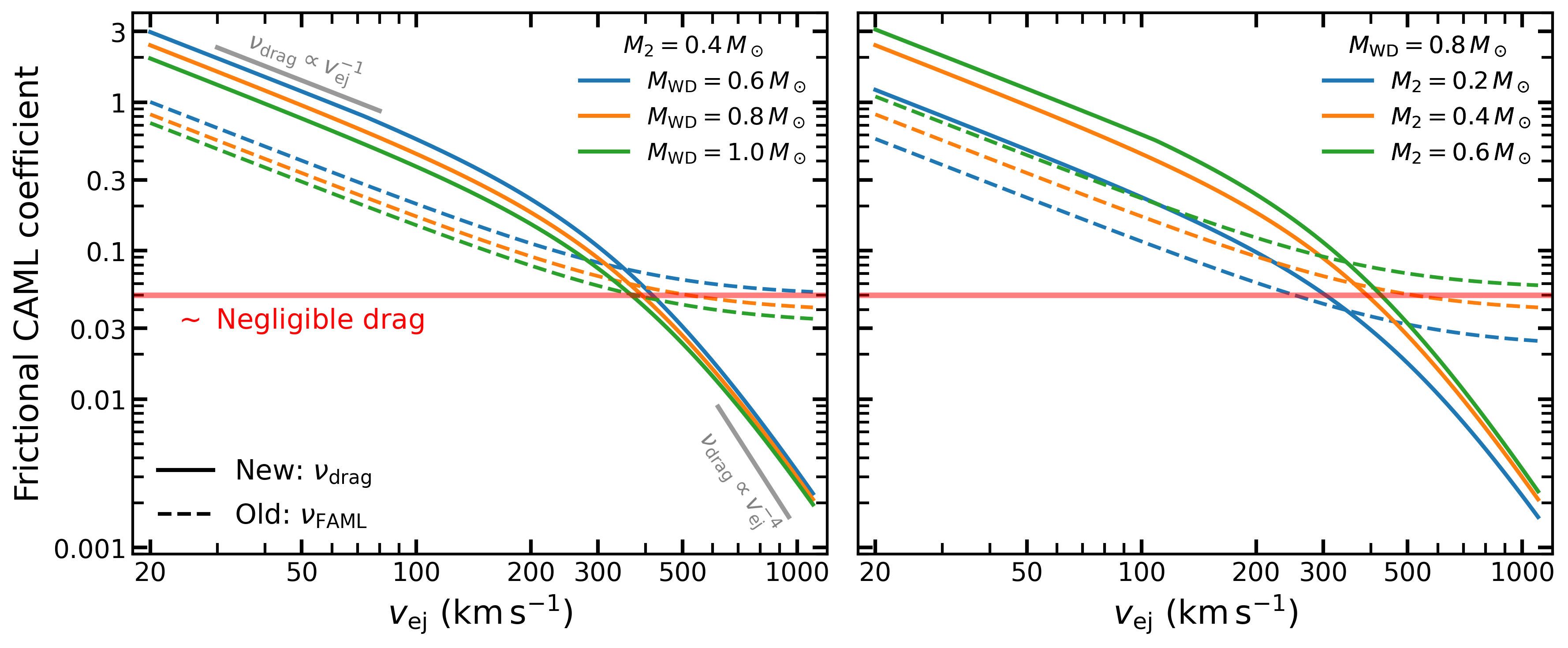}
    \caption{Comparison of the new nova-drag CAML prescription (solid lines) with the FAML prescription of \citet{1998MNRAS.297..633S} (dashed lines) as a function of ejecta velocity. The left panel varies the WD mass at fixed $M_2=0.4\,M_\odot$, while the right panel varies the donor mass at fixed $M_\mathrm{WD} = 0.8\,M_\odot$. By explicitly accounting for the donor's gravitational focusing radius as the interaction scale, our prescription predicts significantly stronger CAML for slow ejecta and weaker CAML for fast ejecta.}
    \label{fig:nudrag_vs_nufaml}
\end{figure*}

\begin{figure}
    \centering
    \includegraphics[width=0.99\linewidth]{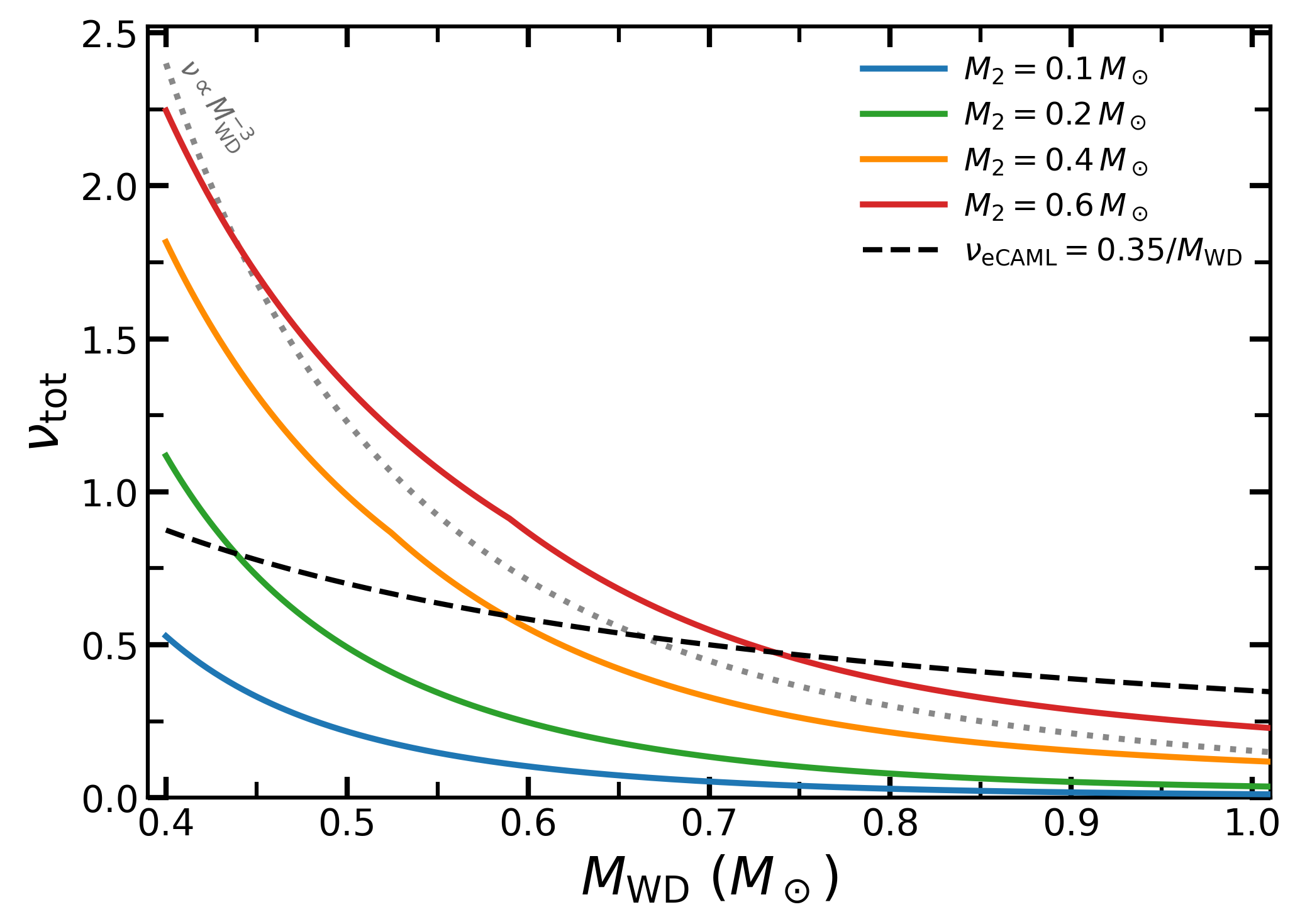}
    \caption{Total CAML coefficient, $\nu_{\rm tot}=\nu_{\rm iso}+\nu_{\rm drag}$, as a function of WD mass for our fiducial model. The coloured solid lines show the results for different donor masses. The black dashed line shows the eCAML prescription for comparison \citep{2016MNRAS.455L..16S}. The nova-drag prescription produces a steeper dependence on WD mass than eCAML and introduces an additional dependence on the donor mass.}
    \label{fig:nu_vs_Mwd}
\end{figure}

To test the derived nova-drag prescription we consider different combinations of $(M_\mathrm{WD},M_2)$, and assume that the donor fills its Roche lobe. The binary separation is therefore fixed by the donor radius and the Roche-lobe relation \citep{1983ApJ...268..368E}. We use a simple mass-radius scaling $R_2/R_\odot=M_2/M_\odot$. Varying this relation shifts the implied orbital periods and orbital velocities but has only a modest effect on the calculated drag coefficient. We compute $\Omega$ from Kepler's law and the donor--ejecta relative velocity ($V$) for a given ejecta velocity ($v_\mathrm{ej}$). We then calculate $\nu_\mathrm{drag}$ from Eq.~\ref{eq:nu_drag_final}, and since the flow in typical CVs is mainly supersonic we use $I\simeq\ln(b_{\max}/r_{\min})$, avoiding the need to specify the exact ejecta temperature and sound speed. We checked a wide range of possible WD and donor masses and found that $I$ has only a moderate effect, varying between $0.7\text{--}1.5$.

The main parameter controlling the strength of the drag is the ejecta velocity ($v_\mathrm{ej}$). Figure~\ref{fig:nudrag_vs_nufaml} shows how $\nu_\mathrm{drag}$ changes with $v_\mathrm{ej}$ for different component masses, and compares it with the FAML prescription of \citet{1998MNRAS.297..633S}. The FAML derivation is similar to ours but is based on a Bondi--Hoyle-type ram-pressure drag force and adopts $R_\mathrm{g}=R_2$. Comparing them gives $\nu_\mathrm{drag}/\nu_\mathrm{FAML} = (I/(1+q)) (R_\mathrm{g}/R_2)^2$. The main difference is therefore that the effective interaction radius $R_\mathrm{g}$ and scale factor $I(\mathcal{M})$ are treated more explicitly in our prescription. Thus, $\nu_\mathrm{drag}$ becomes significantly stronger for slow ejecta with $R_\mathrm{g}>R_2$ and weaker for fast ejecta with $R_\mathrm{g}<R_2$. 

This difference helps explain why \citet{2019ApJ...870...22L} found FAML to be too weak to influence CV evolution significantly. Since $R_\mathrm{g}\propto V^{-2}$, one obtains $\nu_\mathrm{drag}\propto V^{-3}v_\mathrm{ej}^{-1}$. In the limit $v_\mathrm{ej}\ll v_\phi$, one has $V\simeq v_\phi$ and therefore $\nu_\mathrm{drag}\propto v_\mathrm{ej}^{-1}$. In the opposite limit $v_\mathrm{ej}\gg v_\phi$, the drag drops much more steeply, $\nu_\mathrm{drag}\propto v_\mathrm{ej}^{-4}$. Therefore, only sufficiently slow ejecta can produce a substantial drag contribution. For ejecta velocities of $\gtrsim300\,\mathrm{km\,s^{-1}}$, the drag contribution becomes negligible, whereas slow ejecta with $v_\mathrm{ej}\sim 100 \,\mathrm{km\,s^{-1}}$ can readily produce $\nu_\mathrm{drag}\sim1$.

Figure~\ref{fig:nu_vs_Mwd} shows the resulting relation between $\nu_{\rm tot}$ and $M_{\rm WD}$ for our fiducial ejecta-velocity prescription (see Eq.~\ref{eq:v_ej} and Appendix~\ref{app:velocity}) and several representative donor masses. For comparison, we also show the eCAML prescription. Because the ejecta velocity increases steeply with WD mass ($v_\mathrm{ej} \propto M_\mathrm{WD}^3$), the drag contribution decreases more rapidly with $M_{\rm WD}$ than the eCAML coefficient. The nova-drag prescription also introduces an explicit dependence on $M_2$, with more massive donors generally producing stronger CAML because of their larger gravitational focusing radii. Consequently, the drag mechanism preferentially enhances CAML in systems containing low-mass WDs, particularly when they are paired with relatively massive donors.

\section{Effects on CV evolution}
\label{sec:result}

\begin{figure*}
    \centering
    \includegraphics[width=0.7\linewidth]{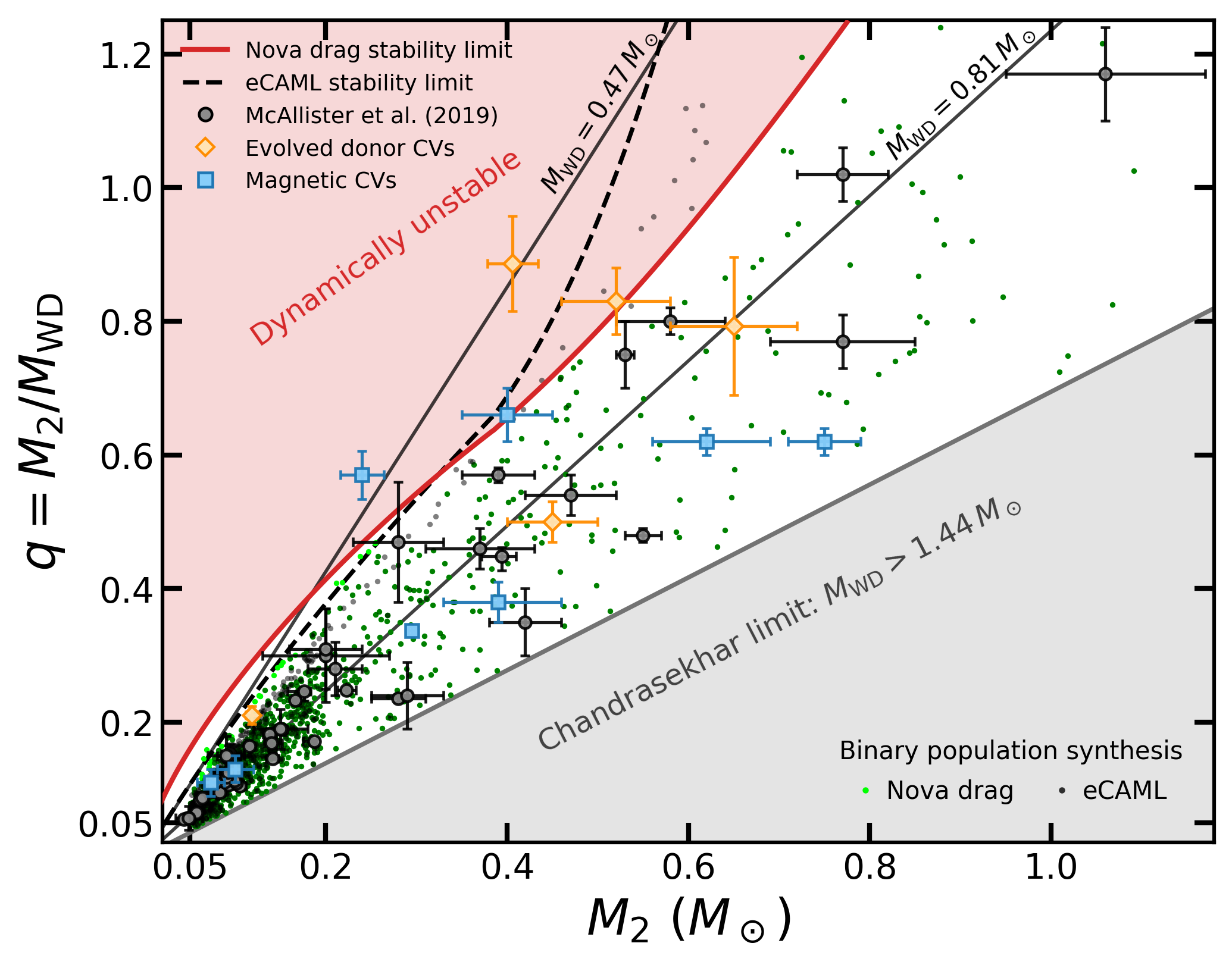}
    \caption{Dynamical stability of mass transfer in the $q$--$M_2$ plane. The red shaded region corresponds to dynamically unstable mass transfer as predicted by our fiducial nova-drag model. The black dashed line shows the stability boundary for eCAML \citep{2016MNRAS.455L..16S}. Small lime and black points represent the surviving nova-drag and eCAML populations predicted by binary population synthesis. Note that most of the points overlap, since the corresponding systems are allowed to survive in both models. Black circles, orange diamonds, and blue squares denote observed ordinary, evolved donor, and magnetic CVs, respectively. The solid black lines indicate constant WD masses of $M_\mathrm{WD}=0.47\,M_\odot$, approximately corresponding to the upper mass limit for He-core WDs, and $M_\mathrm{WD}=0.81\,M_\odot$, corresponding to the mean WD mass observed in CVs.}
    \label{fig:stability}
\end{figure*}

\begin{figure*}
    \centering
    \includegraphics[width=0.99\linewidth]{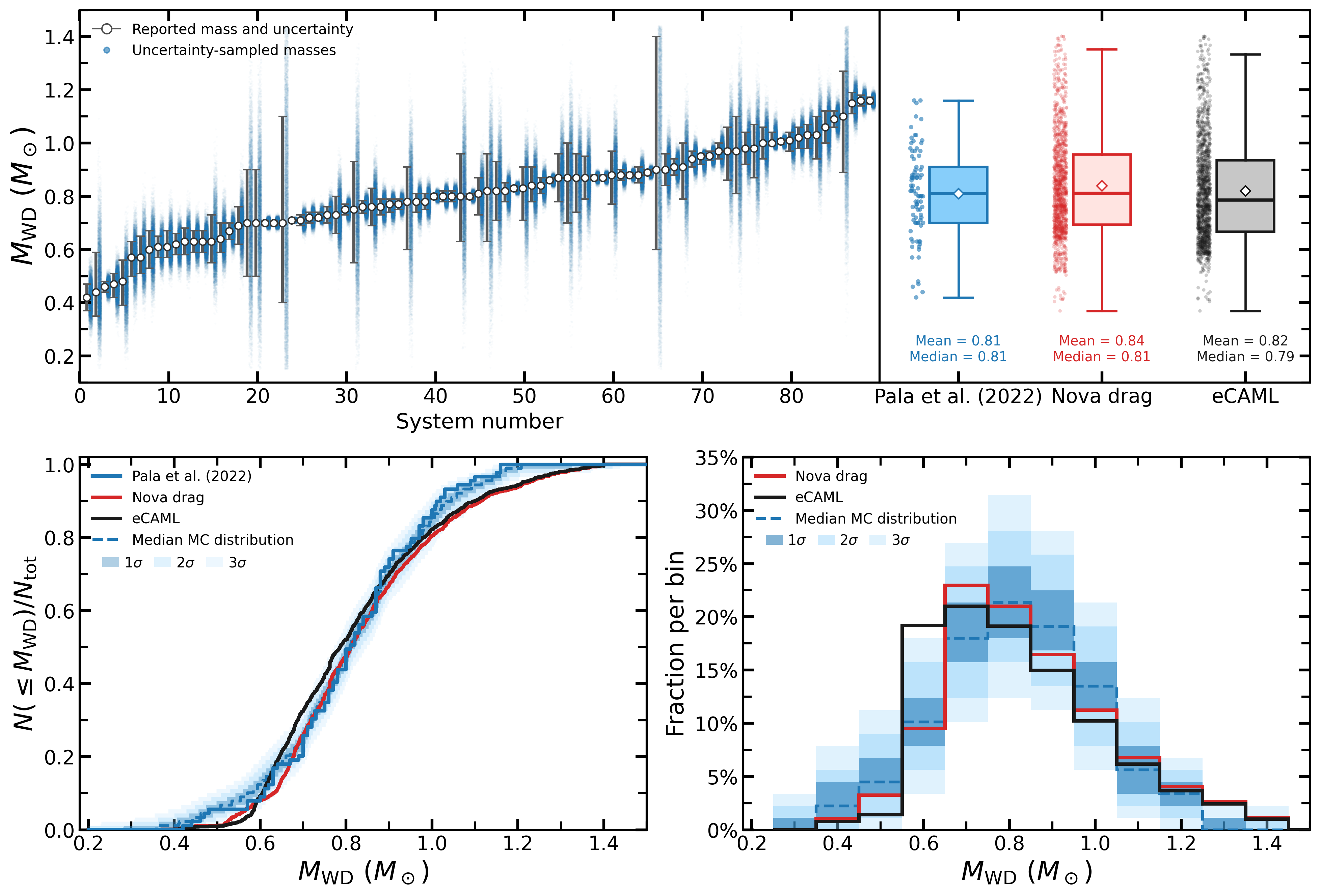}
    \caption{Comparison between the observed and predicted WD mass distributions. Upper left: measured masses and uncertainties for 89 CVs from the observational sample of \citet{2022MNRAS.510.6110P}, with 10\,000 uncertainty-sampled Monte Carlo draws plotted alongside each system. Upper right: box plots of the observed and model populations. Lower panels: cumulative distributions (left) and histograms (right). Dashed blue curves show the median uncertainty-sampled observational distributions and shading indicates the corresponding Gaussian-equivalent $1\sigma$, $2\sigma$, and $3\sigma$ intervals.}
    \label{fig:distributions}
\end{figure*}

We performed binary population synthesis with the Binary Star Evolution code (BSE; \citealt{2002MNRAS.329..897H}), using the version updated and calibrated for CV evolution by \citet{2018MNRAS.478.5626B}. This version of BSE has been successfully adopted in several works \citep[e.g.][]{2020MNRAS.491.5717B, 2022MNRAS.511.4060K, 2025A&A...698L..22S} and reproduces the results from detailed CV evolution calculations with good precision. Upgrades particularly relevant here include revised stability criteria for dynamical and thermal mass transfer, depending on the donor's mass–radius exponents and the assumed CAML. We follow each binary from its primordial configuration through the detached pre-CV and mass-transferring CV phases, evaluating mass-transfer stability at every evolutionary step.

We adopted the initial binary population described in detail in \citet{2024A&A...682A..33B}. In particular, we generated $\approx6\times10^5$ zero-age main-sequence binaries, drawing primary masses between 1--8$\,M_\odot$ from the canonical initial mass function of \citet{2001MNRAS.322..231K}. The corresponding mass ratios and orbital periods were drawn from the correlated distributions of \citet{2017ApJS..230...15M}. We assumed solar metallicity, $Z=0.02$, a constant star formation rate, and a Galactic age of $10\,$Gyr. The binaries were evolved using the standard prescriptions for AML through gravitational wave radiation and magnetic braking \citep{1983ApJ...275..713R} for which we adopted the normalization factors from \citet{2016MNRAS.457.3867Z}. For our fiducial simulations, we used a common-envelope efficiency of $\alpha_\mathrm{CE}=0.25$ \citep{2010A&A...520A..86Z}, the ejecta-velocity prescription $v_{\rm ej}=400 (M_{\rm WD}/0.8\,M_\odot)^3 \,\mathrm{km\,s^{-1}}$ for the nova-drag model, and $C=0.30$ when comparing with the eCAML model. 

Following \citet{2016MNRAS.455L..16S}, we restricted the present-day CV populations to systems with $M_2>0.05\,M_\odot$. Without this cut, the predicted populations are dominated by post-period-minimum systems, whose observed incidence is much lower than predicted by standard population models (the period-bouncer problem; \citealt{2020MNRAS.494.3799P}). One proposed explanation is that the late emergence of a WD magnetic field causes many period bouncers to detach for several gigayears, removing them from the observable accreting CV population \citep{2023A&A...679L...8S, 2025A&A...698L..22S}.

\subsection{Mass transfer stability}

The principal effect of CAML on the predicted CV population arises through its influence on the stability of mass transfer. Mass transfer is dynamically stable when the adiabatic response of the donor exceeds that of its Roche lobe:
\begin{equation}
\label{eq:stability_crit}
\left(\frac{d\ln R_2}{d\ln M_2}\right)_\mathrm{ad}
\equiv \zeta_\mathrm{ad}
> \zeta_\mathrm{RL}
\equiv \frac{d\ln R_{\mathrm{RL},2}}{d\ln M_2}.
\end{equation}
\noindent
For fully non-conservative mass transfer with CAML, the Roche-lobe response is
\begin{equation}
\label{eq:zeta_RL}
\begin{multlined}
\zeta_{\rm RL} =
\frac{2}{3}\left(
\frac{\ln(1+q^{1/3}) - 0.5q^{1/3}/(1+q^{1/3})}
{0.6q^{2/3} + \ln(1+q^{1/3})}
\right) + \\
+ 2\nu_\mathrm{tot} + \frac{q}{1+q} - 2,
\end{multlined}
\end{equation}
\noindent where we have used the Roche-lobe approximation of \citet{1983ApJ...268..368E}. We adopted the prescription for the donor response $\zeta_{\rm ad}(M_2)$ from \citet{1989PhDT.........7H}, which is given by \citep[see also][their equation~A13]{2018MNRAS.478.5626B}:
\begin{equation}
\label{eq:zeta_ad}
\zeta_{\rm ad}(M_2) =
\begin{cases}
-\dfrac{1}{3}, & M_2 \leq 0.38412 M_\odot, \\[1.2ex]
f(M_2), & 0.38412 < M_2/M_\odot \leq 0.8 ,
\end{cases}
\end{equation}
\noindent
where $f(M_2) = 0.782 - 7.464M_2 + 13.925M_2^2 - 5.389M_2^3$. We verified that replacing this prescription with more recent calculations of the donor's adiabatic response \citep[e.g.][]{2015ApJ...812...40G} changes the resulting stability boundaries only marginally and does not affect our conclusions.

Figure~\ref{fig:stability} shows the dynamically stable region predicted by our fiducial nova-drag model in the $q\text{--}M_2$ plane, and compares it to eCAML. The two stability boundaries are remarkably similar over the region occupied by most short-period CVs, despite the substantially different $\nu\text{--}M_\mathrm{WD}$ relations predicted by the two prescriptions. This similarity arises because the stability boundaries depend on where each prescription crosses the critical CAML strength for unstable mass transfer, rather than on the full shape of its $\nu\text{--}M_\mathrm{WD}$ relation. Both models efficiently eliminate systems with low-mass WDs, but the nova-drag model allows a small number of systems containing WDs with masses close to the He-core WD limit ($M_\mathrm{WD}=0.47\,M_\odot$) to survive. This happens only for low-mass donors and is a consequence of the drag becoming weaker for smaller donors (see Figure~\ref{fig:nu_vs_Mwd}). 

Figure~\ref{fig:stability} also shows the surviving populations predicted by our BSE simulations and compares them with observed CVs. Most of the synthetic nova-drag and eCAML populations overlap, since the corresponding systems are allowed to survive in both models. The observational sample is primarily based on \citet{2019MNRAS.486.5535M}, supplemented with several additional systems, particularly evolved-donor and magnetic CVs. Most CVs lie within the dynamically stable region predicted by both prescriptions. 

However, one evolved-donor CV (KIC 5608384; \citealt{2019MNRAS.489.1023Y}) and one magnetic CV (HY Eri; \citealt{2020A&A...634A..91B}) with low-mass WDs lie inside the region predicted to be dynamically unstable. These unusual systems suggest that evolved and magnetic systems may require a more detailed treatment. 

For evolved CVs, the discrepancy may arise because the donor's adiabatic response to mass loss differs from that adopted for unevolved main-sequence donors in our stability calculation. For magnetic CVs, magnetically channelled accretion and a potentially different ejecta geometry can modify the interaction between the nova ejecta and the binary \citep{2011A&A...536A..97F}. In addition, we note that the error bars shown in Figure~\ref{fig:stability} represent $1\sigma$ uncertainties. Given these uncertainties, the apparent disagreement of the two systems with the predicted stability boundary may not be statistically significant.

As a complementary check, we performed detailed binary evolution simulations using the MESA code for representative CVs close to the analytical stability limit shown in Figure~\ref{fig:stability}. Their mass-transfer behaviour agrees closely with the predicted stable and unstable regimes, supporting the adopted stability treatment (see Appendix~\ref{app:mesa}).

\subsection{White dwarf mass distribution}
\label{sec:WDmass}

Figure~\ref{fig:distributions} compares the predicted WD mass distributions with the observational sample of \citet{2022MNRAS.510.6110P}. To describe the effect of measurement uncertainties, we propagated the reported observational mass errors through 10\,000 Monte Carlo realizations, drawing each system's mass using its asymmetric uncertainties. We then constructed a cumulative distribution and histogram for each realization. The dashed curves and shaded bands in the lower panels of Figure~\ref{fig:distributions} show the pointwise medians and corresponding Gaussian-equivalent $1\sigma$, $2\sigma$, and $3\sigma$ intervals, respectively. Observed and model distributions were each normalized to unity, so the statistical comparison tests their shapes.

The nova-drag model reproduces the observed WD mass distribution remarkably well, with the median value of $0.81\,M_\odot$ matching observations. The maximum absolute difference reaches $\approx10\%$ at masses around $M_\mathrm{WD}=1\,M_\odot$. The observational sample contains no system with $M_\mathrm{WD}\gtrsim1.2\,M_\odot$, in contrast to model predictions. This deficit likely reflects observational selection, since \citet{2022MNRAS.510.6110P} excluded novae and nova-like systems from their ultraviolet analysis because the underlying WD could not be reliably isolated. High-mass WDs generally have shorter nova recurrence times ($t_\mathrm{rec} \lesssim 100\,$yr; \citealt{2013ApJ...777..136W}), so excluding novae could contribute to an underrepresentation of massive WDs.

We computed two-sample Kolmogorov--Smirnov (KS) tests between the models and the observed masses, and got $p=0.34$ for the nova-drag model and $p=0.16$ for eCAML. Neither model is therefore rejected at conventional significance levels. To account for the observational uncertainties, we repeated the KS test for the 10\,000 uncertainty Monte Carlo realizations. Only $0.1\%$ and $1.1\%$ of the realizations reject the nova-drag and eCAML populations at the $5\%$ significance level, respectively. The agreement is therefore robust against the measurement uncertainties.

To assess the sensitivity of our results to the model assumptions, we explored the three ejecta velocity prescriptions described in Appendix~\ref{app:velocity}, three eCAML normalizations ($C=0.3$, $0.33$, and $0.35$), and three values of the common-envelope efficiency ($\alpha_\mathrm{CE}=0.20$, $0.25$, and $0.30$). 
The ejecta-velocity prescription is the main driver of the nova-drag results, as it directly controls the drag strength and hence the surviving population. Nevertheless, the adopted prescriptions motivated by nova calculations yield WD mass distributions in good agreement with the observations. At each tested common-envelope efficiency the fiducial nova-drag prescription yields smaller KS distances and lower rejection fractions than any of the tested eCAML normalizations. The complete model grid, distribution plots, and statistical results are presented in Appendix~\ref{app:bps_grid}.

\section{Discussion}
\label{sec:discuss}

The current observational CV WD mass sample does not distinguish between the eCAML and nova-drag prescriptions, which also predict similar numbers of surviving CVs and orbital-period distributions (see Table~\ref{tab:bps_grid}). These results suggest that nova-driven dynamical friction provides the physical mechanism underlying eCAML's success in solving the WD mass problem while preserving the improvements in the predicted orbital-period distribution and space density. A dedicated comparison with the most recent observational samples \citep[e.g.,][]{2024A&A...682L...7S, 2026arXiv260722836I, 2026arXiv260727855H, 2026arXiv260728066S} will be required for further validation. 

Our results form part of a broader reassessment of AML in CVs, including the strength and disruption of magnetic braking \citep{2025A&A...696A..92B, 2026A&A...708L..11B, 2026A&A...710A.317D}. Systemic AML drives mass transfer, while nova-driven CAML provides an additional AML channel coupled to the transferred mass. This feedback can enhance mass transfer in surviving systems or drive it unstable when sufficiently strong. Combining physically motivated prescriptions for magnetic braking and nova-driven CAML will allow their coupled effects on mass transfer and CV evolution to be explored.

The importance of interaction between nova ejecta and the binary is becoming increasingly apparent. Nova calculations suggest that substantial mass loss can be initiated and shaped by the companion before a fast wind develops \citep{2022ApJ...938...31S}, while recent three-dimensional simulations show that the ejecta acquire additional angular momentum through binary interaction \citep{2026PhRvD.113l3023N}. Such interaction may also have observable consequences beyond outflow dynamics and orbital evolution. The ejecta can disrupt the accretion disk and chemically contaminate the companion producing abundance anomalies observed in some CVs \citep{2012ApJ...745...43N, 2018A&A...613A...8F, 2021ApJ...914....5S}.

Orbital period jumps across nova eruptions provide a direct observational test of our proposed drag mechanism. Assuming an ejected mass $\Delta M_\mathrm{ej}=-\Delta M_\mathrm{WD}>0$, negligible donor mass change, and negligible systemic AML during the eruption, the fractional period change is
\begin{equation}
\label{Pdot_nova}
    \frac{\Delta P}{P} \simeq \frac{\Delta M_\mathrm{ej}}{M_2} \left(\frac{2q}{1+q} - 3\nu_\mathrm{drag}\right).
\end{equation}
\noindent Here, we retain only terms linear in the fractional mass and angular-momentum changes, which is an accurate approximation for our case. 

\citet{2023MNRAS.525..785S} reported both positive and negative orbital-period changes across nova eruptions, with the latter being difficult to reconcile with standard mass-loss prescriptions. Our model naturally allows for both signs, depending on the competition between orbital expansion caused by mass ejection and orbital contraction driven by dynamical friction. For the adopted ejecta-velocity prescriptions, the sign is primarily determined by the WD mass, with a transition at approximately $M_{\rm WD}=0.6\text{--}0.7\,M_\odot$. Systems containing less massive WDs are therefore predicted to experience negative period jumps, while positive jumps are expected for more massive WDs.

Interestingly, this prediction qualitatively resembles the distinction between classical and recurrent novae reported by \citet{2023MNRAS.525..785S}. However, the observational evidence should be treated with considerable caution since measuring the small orbital-period changes associated with nova eruptions is intrinsically challenging, particularly when pre-eruption periods are inferred from sparse archival photographic observations spanning long time intervals. A robust test will require accurately determined orbital periods before and after nova eruptions, together with independent constraints on the WD mass, ejecta mass, and ejecta velocity.

Our results corroborate that nova eruptions lead to unstable mass transfer in CVs containing low-mass WDs. This should result in merger events, while producing observable transients. \citet{2021ApJ...923..100M} predicted that CV merger transients resemble slow classical novae and luminous red novae from ordinary stellar mergers. Following the merger, residual hydrogen shell burning can inflate the remnant into a giant for thousands of years, masking a population of CV merger remnants among Galactic giants \citep{2021ApJ...923..100M}. Searches for traces of these events could reveal the fate of the missing low-mass WD CVs and test the connection between nova eruptions and binary survival. After the envelope is lost or exhausted, single low-mass He-core WDs can form, providing a formation channel capable of accounting for their observed abundance \citep{2017MNRAS.466L..63Z, 2018MNRAS.474..427Z}. Nova-driven mergers may therefore connect the scarcity of low-mass WDs in CVs with their otherwise puzzling presence among isolated WDs.

\section{Conclusion}
\label{sec:conclude}

We developed a new model for consequential angular momentum loss, the nova-drag model, which explains the origin of the enhanced angular momentum loss required in CV evolution. The key link is provided by nova ejecta and gaseous dynamical friction: slower outflows from lower-mass WDs are more strongly gravitationally focused by the companion, producing enhanced dynamical drag and thereby extracting more orbital angular momentum. This preferentially drives systems with low-mass WDs into dynamically unstable mass transfer. Our CV population synthesis yields surviving CV populations that reproduce the observed WD mass distribution of CVs. They also closely resemble those predicted by the empirical eCAML prescription, which has been remarkably successful in reproducing the observed properties of the CV population. This indicates that a single, unified physical process, the dynamical friction of nova ejecta on the companion, can resolve the major discrepancies that have defined CV research for the past decade and a half. More broadly, our results show that brief, violent nova eruptions leave a permanent imprint on the binary population, sculpting the mass distribution of an entire class of compact binaries. Since consequential angular momentum loss determines the survival of interacting white dwarf binaries, the mechanism may also affect evolutionary pathways leading to Type Ia supernova progenitors.

\begin{acknowledgments}
VD acknowledges support from Nanjing University. XDL was supported by the National Natural Science Foundation of China under grant Nos. 12121003 and 12633010. This work is supported by the National Natural Science Foundation of China (NSFC, Nos. W2632018, 12288102, 12125303, 12525304), the Strategic Priority Research Program of the Chinese Academy of Sciences (grant Nos. XDB1160303, XDB1160300, XDB1160000, XDB1160200, XDB1160201), the CAS Project for Young Scientists in Basic Research (YSBR-148), the Yunnan Revitalization Talent Support Program ``YunLing Scholar'' project, International Centre of Supernovae (ICESUN), Yunnan Key Laboratory of Supernova Research (No. 202505AV340004), the New Cornerstone Science Foundation through the XPLORER PRIZE, and the Yunnan Revitalization Talent Support Program-Science \& Technology Champion Project (No. 202305AB350003). MRS thanks for support from the research unit FOR2990 (eRO-STEP: Stellar endpoints with eROSITA) funded by the Deutsche Forschungsgemeinschaft and the internal research project PI\_LIR\_26\_07 at USM. The authors thank the anonymous referee for useful suggestions that helped improve the paper.
\end{acknowledgments}

\appendix 

\section{Nova ejecta velocities and adopted prescriptions}
\label{app:velocity}

A systematic increase in nova ejecta velocity with WD mass is expected because more massive WDs have deeper gravitational potential wells. A characteristic velocity scale is provided by the escape velocity:

\begin{equation}
   v_{\rm esc}=\left(\frac{2GM_{\rm WD}}{R_{\rm WD}}\right)^{1/2}. 
\end{equation}

Using the zero-temperature WD mass-radius relation of \citet{1972ApJ...175..417N}, the local dependence around \(M_{\rm WD}=0.8\,M_\odot\) is approximately \(v_{\rm esc}\propto M_{\rm WD}\). However, the ejecta velocity also depends on the eruption dynamics, including the ignition-envelope mass, thermonuclear energy release, and efficiency with which this energy is converted into ejecta kinetic energy. Consequently, its dependence on WD mass can be considerably steeper than that expected from the escape velocity alone.

Because nova ejecta velocities depend on several parameters, including the WD mass, accretion rate, WD core temperature, envelope composition, and outflow phase, a simple analytical prescription cannot capture all the relevant physics. We therefore adopt approximate relations motivated by the trends found in detailed nova calculations \citep{1995ApJ...445..789P, 2005ApJ...623..398Y}.

\begin{figure}
    \centering
    \includegraphics[width=0.495\linewidth]{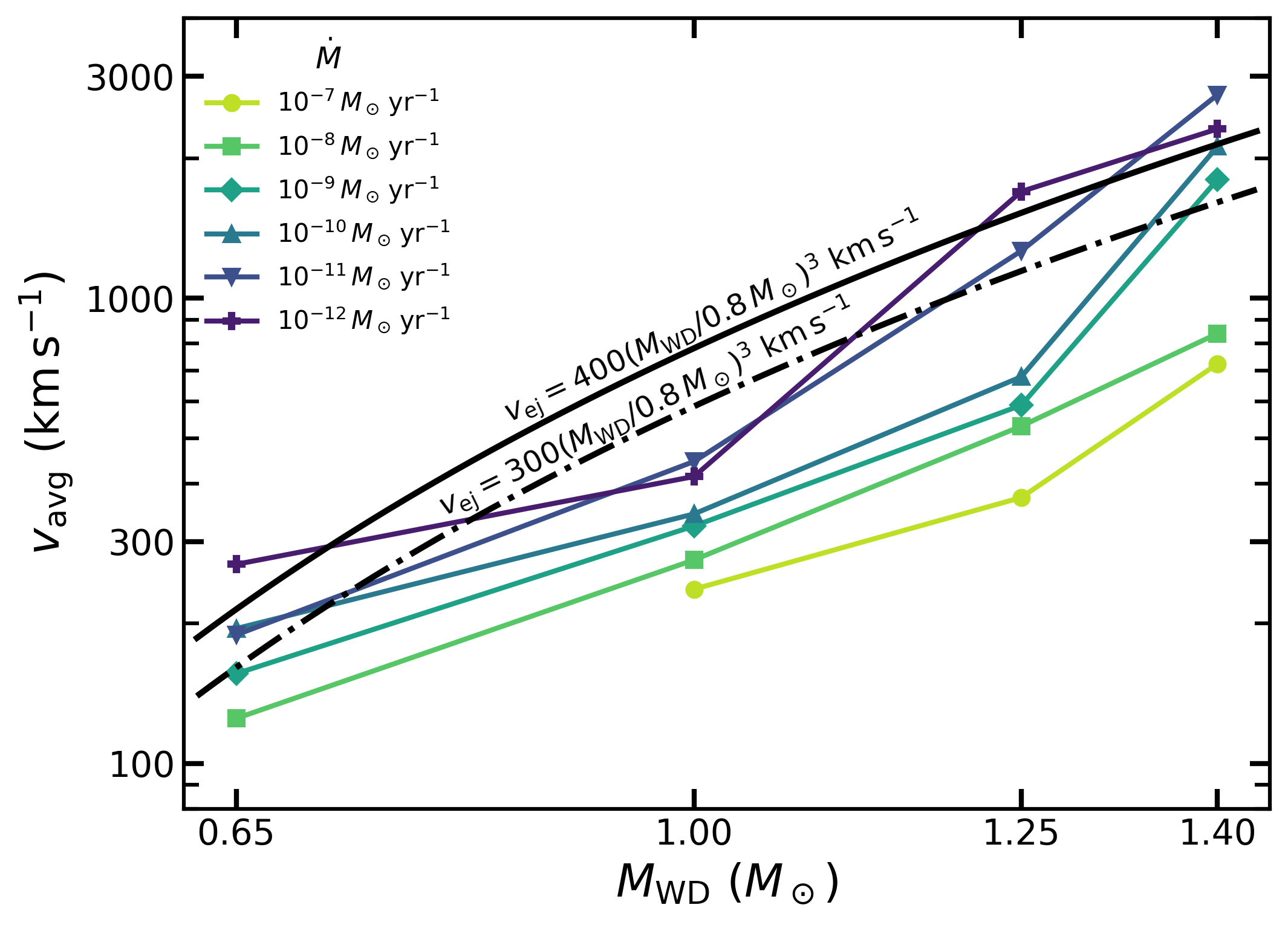}
    \includegraphics[width=0.495\linewidth]{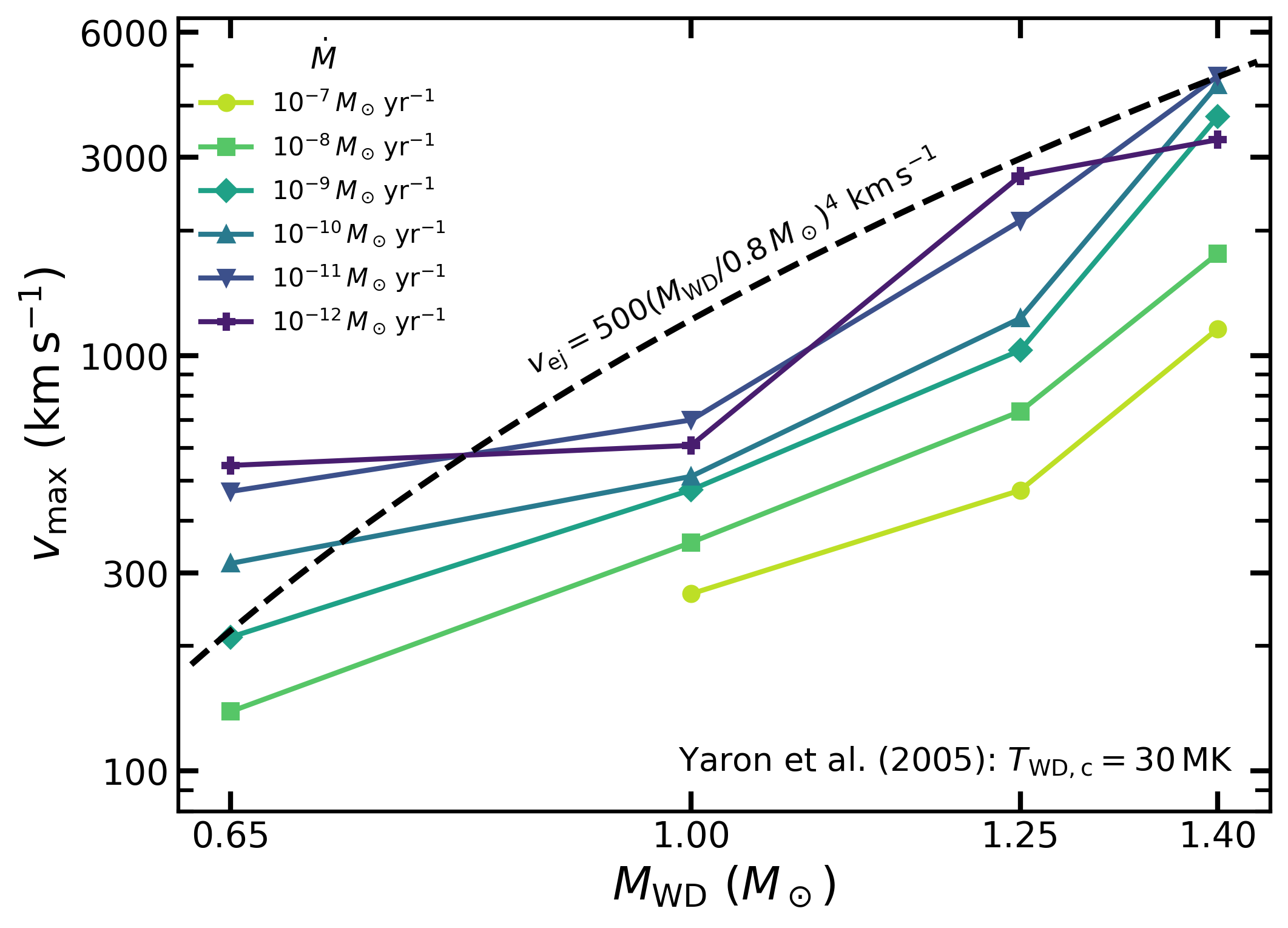}
    \caption{Average (left) and maximum (right) ejecta velocities over the nova mass-loss phase as a function of WD mass for different accretion rates at $T_{\mathrm{WD,\,c}}=30\,\mathrm{MK}$ from the model grid of \citet{2005ApJ...623..398Y}. The black lines show our adopted velocity prescriptions.}
    \label{fig:vej_Yaron}
\end{figure}

Figure~\ref{fig:vej_Yaron} shows the average and maximum ejecta velocities over the nova mass-loss phase as a function of WD mass for different accretion rates, based on the nova-model grid of 
\citet{2005ApJ...623..398Y} for a fixed WD core temperature of \(T_{\rm WD,c}=30\,\mathrm{MK}\). Despite substantial variations with accretion rate, both the average and maximum velocities increase systematically with WD mass, with a dependence considerably steeper than that of the escape velocity.

Motivated by these calculations, we approximate the WD-mass dependence of the ejecta velocity by a power law:

\begin{equation}
    v_{\rm ej}=v_0\left(\frac{M_{\rm WD}}{0.8\,M_\odot}\right)^p,
\end{equation}

\noindent where \(v_0\) is the characteristic ejecta velocity at \(M_{\rm WD}=0.8\,M_\odot\), and \(p\) describes its dependence on WD mass. We consider three prescriptions:

\begin{equation}
    \begin{aligned}
    v_{\rm ej}&=300\left(\frac{M_{\rm WD}}{0.8\,M_\odot}\right)^3\,\mathrm{km\,s^{-1}},\\
    v_{\rm ej}&=400\left(\frac{M_{\rm WD}}{0.8\,M_\odot}\right)^3\,\mathrm{km\,s^{-1}},\\
    v_{\rm ej}&=500\left(\frac{M_{\rm WD}}{0.8\,M_\odot}\right)^4\,\mathrm{km\,s^{-1}}.
    \end{aligned}
\end{equation}

The two cubic relations provide approximate representations of the average velocities shown in the left panel of Figure~\ref{fig:vej_Yaron}, whereas the steeper relation more closely follows the WD-mass dependence of the maximum velocities shown in the right panel. We adopt the intermediate relation, with \(v_0=400\,\mathrm{km\,s^{-1}}\) and \(p=3\), as our fiducial prescription, representing the characteristic expansion velocities averaged over the nova mass-loss phase. The alternative prescriptions allow us to assess the sensitivity of our results to uncertainties in both the normalization and the WD-mass dependence of the ejecta velocity (Appendix~\ref{app:bps_grid}).

Across the WD-mass range \(0.4\)--\(1.4\,M_\odot\), our fiducial prescription yields velocities of approximately \(50\)--\(2100\,\mathrm{km\,s^{-1}}\), broadly spanning the range of slow and fast ejecta components inferred from observations \citep[e.g.][]{2011MNRAS.410..525M, 2012A&A...543A..86N, 2016ApJ...833..149D}. 

However, nova ejecta are not characterized by a single velocity. Time-resolved spectroscopy reveals multiple outflow components, often consisting of an initially slow outflow followed by a faster wind \citep{2020ApJ...905...62A, 2026NatAs..10..271A}. Theoretical calculations similarly suggest that much of the envelope may initially be expelled through a slow, binary-driven outflow before the development of a fast optically thick wind \citep{1994ApJ...437..802K, 2022ApJ...938...31S}. 

For the dynamical-friction mechanism considered here, the relevant quantity is the characteristic velocity of the ejecta interacting with the donor, rather than necessarily the terminal velocity inferred from observations. Moreover, because the drag depends nonlinearly on ejecta velocity, a single characteristic velocity cannot fully describe the contributions of different outflow components. Our adopted prescriptions should therefore be regarded as first-order approximations to the systematic WD-mass dependence suggested by nova calculations. A more realistic treatment will require hydrodynamical calculations accounting for the velocity distribution, ejecta geometry, and fraction of the ejected material participating in the drag interaction.

\section{Validity and limitations of the analytical model}
\label{app:assume}

Our analytical treatment relies on several simplifying assumptions, of which we discuss the most important below (see also \citealt{2024ApJ...974..216O, 2026ApJ..1000..212O} for a broader discussion of the  limitations of semi-analytical treatments of gaseous dynamical friction). 

The finite spatial extent and density structure of the nova ejecta may affect the gravitational wake. However, recent calculations for circular orbits in spherically stratified gas find that the azimuthal drag component responsible for AML remains close to the homogeneous-medium prediction in the supersonic regime \citep{2026arXiv260716422G}. Although these calculations do not directly describe expanding nova ejecta, they support using the homogeneous-medium prescription as a first approximation.

In the slow-outflow regime, the gravitational focusing radius can become comparable to the binary separation, so density gradients across the wake may modify its structure and the effective value of $I$. This concern is most pronounced in systems containing low-mass WDs and relatively massive donors, particularly for $M_\mathrm{WD} \lesssim 0.5\,M_\odot$ and $M_2 \gtrsim 0.5\,M_\odot$, where $R_\mathrm{g} \sim a$. Since these systems already lie well within the dynamically unstable mass-transfer regime in our model, moderate corrections to the drag strength are unlikely to substantially alter the predicted surviving CV population.

Although derived within linear perturbation theory, the \citet{1999ApJ...513..252O} prescription has been shown by hydrodynamical simulations to remain a good estimate in nonlinear supersonic flows when an appropriate inner cutoff ($r_\mathrm{min}$) is adopted \citep{2013ApJ...775...72B, 2025MNRAS.541..396M}. Idealized linear models accounting for both binary components have also reproduced the drag torque during the slow-inspiral phase of a common-envelope simulation when supplied with suitable gas properties \citep{2008ApJ...679L..33K, 2026PASA...43....2B}.

Gas capture by the donor may also modify the momentum exchange. \citet{2024ApJ...966....7S} found that the net drag on accreting objects deviates from the linear prediction in subsonic and mildly supersonic flows, whereas agreement improves at higher Mach numbers when an appropriate inner cutoff is adopted. The strongly supersonic donor--ejecta motion considered here therefore supports treating these effects within an effective cutoff approximation.

A further contribution may arise from direct hydrodynamical drag on the donor. Approximating this force as $F_\mathrm{ram}\approx1/2\pi R_2^2\rho V^2$ and following the same cycle-averaging procedure as in our derivation gives the comparison $\nu_\mathrm{ram}/\nu_\mathrm{drag}\approx1/2(R_2/R_\mathrm{g})^2$. Here we ignored $I$ and the uncertain scale factor $C_\mathrm{drag}$. This estimate is closely related to the FAML prescription of \citet{1998MNRAS.297..633S}. Since $R_\mathrm{g}>R_2$ in the low-velocity regime of interest, direct ram-pressure drag is expected to be subdominant, although it may modestly modify the total CAML (see also Figure~\ref{fig:nudrag_vs_nufaml}).

Finally, our treatment assumes a spherically symmetric ejecta density, whereas nova outflows may be anisotropic and structured. Concentration of the ejecta towards the orbital plane would increase the density encountered by the donor and enhance the drag, whereas concentration away from this plane would reduce it (this may be the case for magnetic CVs; \citealt{2011A&A...536A..97F}). The equatorial concentration of nova ejecta may develop later through interaction with the binary during envelope expansion \citep{2021ARA&A..59..391C}. An initially approximately spherical outflow therefore provides a reasonable starting approximation for calculating the interaction with the donor.

\section{Additional binary evolution and population synthesis calculations}
\label{app:add_simulations}

\subsection{Detailed evolutionary sequences with MESA}
\label{app:mesa}

\begin{figure*}
    \centering
    \includegraphics[width=0.99\linewidth]{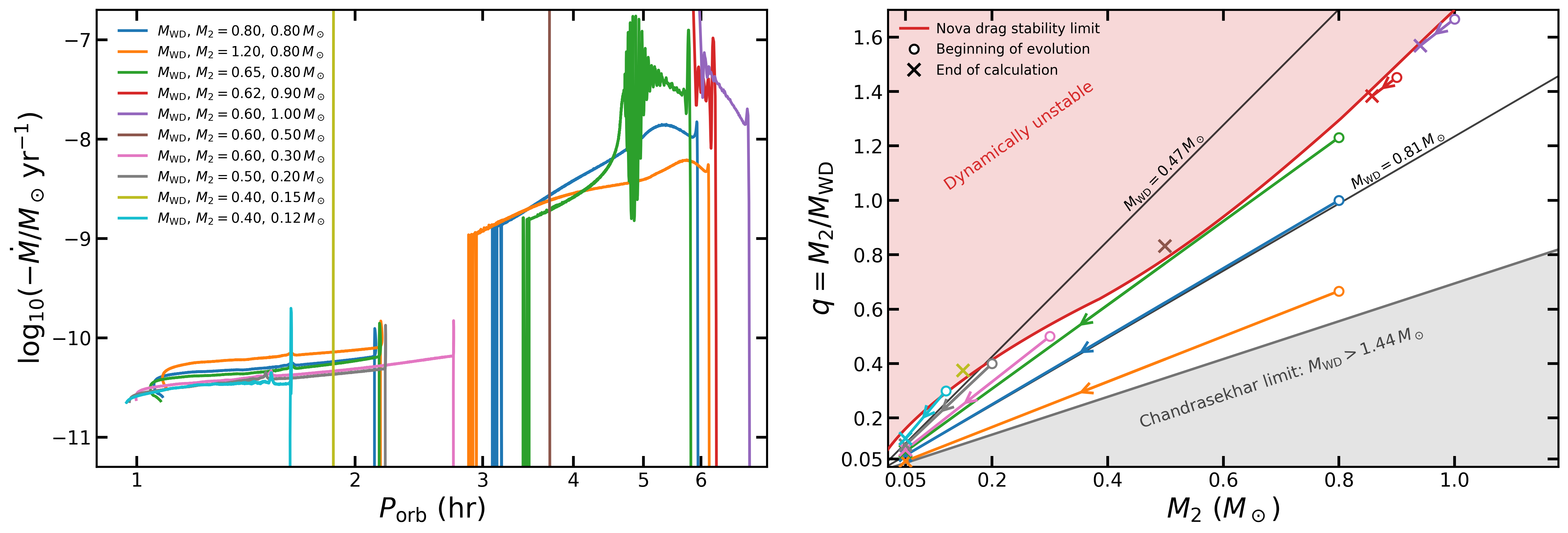}
    \caption{Representative MESA evolutionary sequences calculated with the fiducial nova-drag prescription for different initial WD and donor masses. Left: mass-transfer rate as a function of orbital period. Right: comparison with the analytical stability boundary in the $q\text{--}M_2$ plane. Open circles mark the beginning of evolution, crosses mark the last calculated models, and arrows indicate the direction of evolution.}
    \label{fig:mesa_evolve}
\end{figure*}

To assess the stability predictions using detailed donor models, we implemented the nova-drag CAML prescription in the Modules for Experiments in Stellar Astrophysics (MESA; \citealt{2011ApJS..192....3P, 2013ApJS..208....4P, 2015ApJS..220...15P, 2018ApJS..234...34P, 2019ApJS..243...10P}; \citealt{2023ApJS..265...15J}) code and calculated representative CV evolutionary sequences. These calculations explicitly follow the donor's internal structure and its response to mass loss, providing a test of the stability criterion based on the approximate adiabatic response $\zeta_\mathrm{ad}(M_2)$ from \citet{1989PhDT.........7H} adopted in our BSE population synthesis.

We used MESA version r24.08.1, treating the WD as a point mass and assuming fully non-conservative mass transfer. The binaries were initialized with an orbital period of $P_\mathrm{orb}=1\,$d, and mass transfer was calculated using the \citet{1988A&A...202...93R} prescription. The donor metallicity was fixed at the solar value of $Z=0.02$, and the simulation was terminated when the donor reached a mass of $0.05\,M_\odot$ or when unstable mass transfer began (i.e. $|\dot{M}| \gtrsim 10^{-5}\,M_\odot\,\text{yr}^{-1}$). We selected initial WD and donor masses spanning both sides of the analytical stability boundary, focusing on systems closer to it. Systemic AML was included through standard gravitational wave radiation and magnetic braking \citep{1983ApJ...275..713R}. We adopted our fiducial nova-drag prescription (see Figure~\ref{fig:nu_vs_Mwd}).

Figure~\ref{fig:mesa_evolve} shows the resulting mass-transfer evolutionary tracks and the corresponding trajectories in the $q\text{--}M_2$ plane. The detailed MESA calculations agree closely with the stability predictions adopted in BSE: systems that remain on the stable side of the analytical boundary sustain secular mass transfer, whereas those entering the unstable region exhibit a rapid increase in the mass-transfer rate. Notably, the run with initial $M_\mathrm{WD}, \, M_2=0.65, \, 0.80 \, M_\odot$ approaches the analytical stability boundary closely but maintains stable evolution after episodes of variable mass transfer. The calculations also show that systems containing low-mass WDs close to the He-core limit ($M_\mathrm{WD}=0.40 \text{--} 0.50\,M_\odot$) can survive when paired with very low-mass donors, entering the CV phase below the period gap.

These representative calculations support the use of the approximate donor response adopted in our BSE population synthesis. Reproducing the observed period gap and minimum requires further refinement of the magnetic
braking prescription \citep[e.g.,][]{2025A&A...696A..92B, 2026A&A...708L..11B, 2026A&A...710A.317D}.

\subsection{Sensitivity of the population synthesis to model parameters}
\label{app:bps_grid}

\begin{figure}
    \centering
    \includegraphics[width=0.99\linewidth]{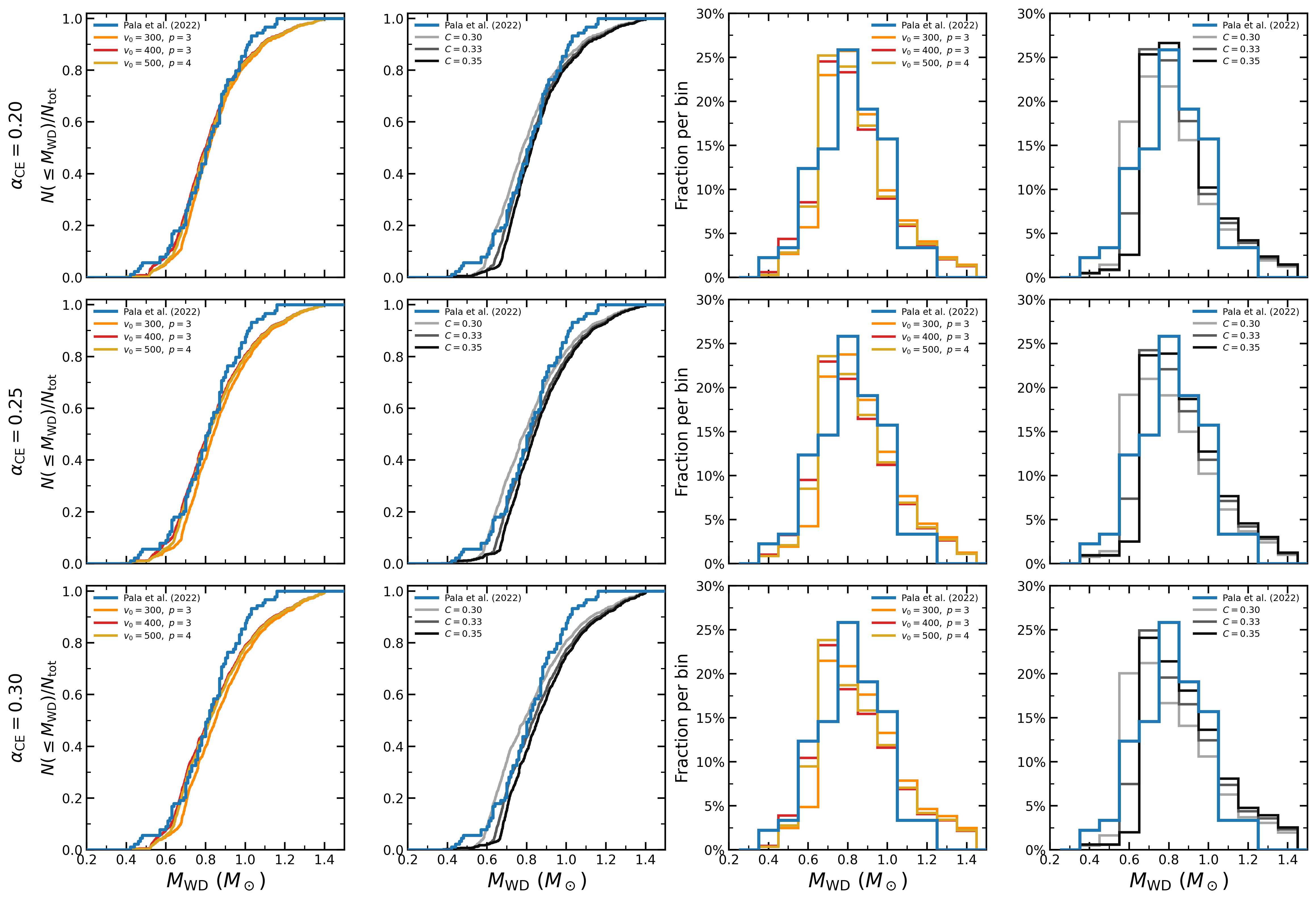}
    \caption{White dwarf mass distributions predicted by the complete grid of binary-population models compared with the observational sample of \citet{2022MNRAS.510.6110P} (blue). The rows correspond to $\alpha_{\rm CE}=0.20$, $0.25$, and $0.30$. The first and second columns show the cumulative distributions predicted by the nova-drag and eCAML models, while the third and fourth columns show the corresponding normalized histograms.}
    \label{fig:allsimulations}
\end{figure}

To examine the sensitivity of our results to the adopted model parameters, we calculated a grid of 18 binary populations. For the nova-drag prescription, we considered the three ejecta-velocity relations from Appendix~\ref{app:velocity}. For comparison, we calculated eCAML populations with $C=0.30$, $0.33$, and $0.35$. Each prescription was evaluated using common-envelope efficiencies of $\alpha_{\rm CE}=0.20$, $0.25$, and $0.30$. The final populations were restricted to systems with $M_2>0.05\,M_\odot$. Table~\ref{tab:bps_grid} reports the number of surviving CVs and the fraction below the period gap, which ranges from $69.1\%$ to $72.6\%$ across the grid. The remarkably similar population sizes and below-gap fractions suggest comparable space densities and orbital-period distributions for the two CAML prescriptions.

Figure~\ref{fig:allsimulations} shows that the predicted WD mass distributions vary systematically but moderately with the adopted parameters. Stronger CAML preferentially removes systems containing lower-mass WDs and shifts the surviving distribution towards higher masses. This trend is particularly clear for eCAML, for which increasing $C$ decreases the number of surviving systems and increases both the mean and median WD masses (see Table~\ref{tab:bps_grid}). A similar effect is found when varying the velocity normalization for the nova-drag model. Increasing $\alpha_{\rm CE}$ also shifts the distributions moderately towards higher WD masses.

To quantify the agreement with observations we performed two-sample Kolmogorov-Smirnov tests with the \texttt{scipy.stats.ks\_2samp} tool. In addition to testing the central measured masses from observations, we repeated the tests for 10\,000 Monte Carlo realizations generated from the observational uncertainties \citep{2022MNRAS.510.6110P}. The statistical results are summarized in Table~\ref{tab:bps_grid}. Among the nova-drag prescriptions, the model with $v_0=400\,{\rm km\,s^{-1}}$ and $p=3$ provides the most consistently good agreement as only $<0.3\%$ of the observational uncertainty realizations reject it. The $v_0=500\,{\rm km\,s^{-1}},\,p=4$ model also performs well, with rejection fractions below $2\%$. The $v_0=300\,{\rm km\,s^{-1}},\,p=3$ model is rejected more often in $\sim25\%$ of cases. For eCAML, $C=0.30$ gives the best overall agreement and is rejected in $<5\%$ of cases, whereas larger values of $C$ progressively overproduce high-mass WDs. The $C=0.35$ populations are rejected in $\sim75\%$ of the uncertainty realizations. These results demonstrate that the agreement between the nova-drag populations and the observations is remarkably good, and not confined to a single finely tuned set of parameters.

\begin{table}
\caption{Comparison of the CV population synthesis grid with the observed WD mass distribution.}
\label{tab:bps_grid}
\renewcommand{\arraystretch}{1.15}
\setlength{\tabcolsep}{4.0pt}
\noindent
\begin{tabular}{c l r r r r r r c r}
\hline
\noalign{\vskip 2pt}
$\alpha_{\rm CE}$ &
Model &
$N_\mathrm{CV}$ &
\shortstack{$f_{\rm below}$\\($\%$)} &
\shortstack{$\langle M_{\rm WD}\rangle$\\$(M_\odot)$} &
\shortstack{$\widetilde{M}_{\rm WD}$\\$(M_\odot)$} &
$D_{\rm KS}$ &
$p_{\rm KS}$ &
\shortstack{$D_{\rm KS}^{\rm MC}$\\($1\sigma$ interval)} &
\shortstack{$f_{\rm rej}$\\($\%$)}
\\
\hline

0.20 & Nova drag: $v_0=300,\ p=3$
& 1053 & 70.8 & 0.850 & 0.820 & 0.099 & 0.370
& $[0.100,\,0.143]$ & 11.5 \\

0.20 & Nova drag: $v_0=400,\ p=3$
& 1162 & 70.9 & 0.827 & 0.801 & 0.074 & 0.730
& $[0.070,\,0.099]$ & 0.1 \\

0.20 & Nova drag: $v_0=500,\ p=4$
& 1131 & 71.0 & 0.835 & 0.807 & 0.088 & 0.514
& $[0.078,\,0.116]$ & 1.0 \\

0.20 & eCAML: $C=0.30$
& 1249 & 70.9 & 0.815 & 0.779 & 0.098 & 0.376
& $[0.086,\,0.121]$ & 1.4 \\

0.20 & eCAML: $C=0.33$
& 1099 & 70.0 & 0.844 & 0.811 & 0.123 & 0.151
& $[0.102,\,0.146]$ & 15.1 \\

0.20 & eCAML: $C=0.35$
& 1018 & 70.1 & 0.860 & 0.826 & 0.143 & 0.062
& $[0.140,\,0.185]$ & 72.5 \\

\hline

0.25 & Nova drag: $v_0=300,\ p=3$
& 1031 & 69.3 & 0.869 & 0.841 & 0.124 & 0.149
& $[0.120,\,0.160]$ & 32.9 \\

0.25 & Nova drag: $v_0=400,\ p=3$
& 1167 & 70.4 & 0.839 & 0.812 & 0.101 & 0.338
& $[0.075,\,0.105]$ & 0.1 \\

0.25 & Nova drag: $v_0=500,\ p=4$
& 1137 & 70.4 & 0.847 & 0.816 & 0.106 & 0.291
& $[0.085,\,0.122]$ & 1.7 \\

0.25 & eCAML: $C=0.30$
& 1281 & 70.8 & 0.821 & 0.786 & 0.122 & 0.156
& $[0.084,\,0.118]$ & 1.1 \\

0.25 & eCAML: $C=0.33$
& 1109 & 70.2 & 0.855 & 0.824 & 0.124 & 0.143
& $[0.106,\,0.148]$ & 16.1 \\

0.25 & eCAML: $C=0.35$
& 1027 & 69.1 & 0.872 & 0.842 & 0.140 & 0.073
& $[0.136,\,0.180]$ & 68.0 \\

\hline

0.30 & Nova drag: $v_0=300,\ p=3$
& 963 & 70.9 & 0.880 & 0.848 & 0.149 & 0.049
& $[0.125,\,0.161]$ & 36.9 \\

0.30 & Nova drag: $v_0=400,\ p=3$
& 1100 & 71.5 & 0.848 & 0.815 & 0.122 & 0.160
& $[0.087,\,0.117]$ & 0.3 \\

0.30 & Nova drag: $v_0=500,\ p=4$
& 1073 & 71.6 & 0.855 & 0.820 & 0.126 & 0.132
& $[0.094,\,0.126]$ & 1.7 \\

0.30 & eCAML: $C=0.30$
& 1205 & 72.6 & 0.828 & 0.789 & 0.137 & 0.080
& $[0.095,\,0.131]$ & 4.1 \\

0.30 & eCAML: $C=0.33$
& 1026 & 71.2 & 0.868 & 0.833 & 0.135 & 0.090
& $[0.119,\,0.159]$ & 32.1 \\

0.30 & eCAML: $C=0.35$
& 938 & 70.7 & 0.889 & 0.856 & 0.159 & 0.030
& $[0.151,\,0.192]$ & 87.6 \\

\hline
\end{tabular}

\tablecomments{$N_\mathrm{CV}$ is the number of surviving simulated systems with $M_2>0.05\,M_\odot$. $f_{\rm below}$ is the percentage of systems below the period gap ($P_{\rm orb}<2\,\mathrm{hr}$). $\langle M_{\rm WD}\rangle$ and $\widetilde{M}_{\rm WD}$ are the mean and median WD masses, respectively. $D_{\rm KS}$ and $p_{\rm KS}$ are the two-sample Kolmogorov--Smirnov statistic and its associated p-value when comparing the modelled WD mass distribution with the observational sample from \citet{2022MNRAS.510.6110P}. $D_{\rm KS}^{\rm MC}$ gives the 16th--84th percentile interval of the statistic obtained from 10\,000 Monte Carlo realizations of the observational uncertainties. $f_{\rm rej}$ is the percentage of realizations for which the model is rejected at the $5\%$ significance level ($p_{\rm KS}<0.05$).}

\end{table}

\bibliography{NovadraginCVs}{}

@BOOK{2003cvs..book.....W,
       author = {{Warner}, Brian},
        title = "{Cataclysmic Variable Stars}",
         year = 2003,
          doi = {10.1017/CBO9780511586491},
       adsurl = {https://ui.adsabs.harvard.edu/abs/2003cvs..book.....W}
}

@ARTICLE{2011ApJS..194...28K,
       author = {{Knigge}, Christian and {Baraffe}, Isabelle and {Patterson}, Joseph},
        title = "{The Evolution of Cataclysmic Variables as Revealed by Their Donor Stars}",
      journal = {\apjs},
         year = 2011,
        month = jun,
       volume = {194},
       number = {2},
          eid = {28},
        pages = {28},
          doi = {10.1088/0067-0049/194/2/28},
archivePrefix = {arXiv},
       eprint = {1102.2440},
 primaryClass = {astro-ph.SR},
       adsurl = {https://ui.adsabs.harvard.edu/abs/2011ApJS..194...28K}
}

@ARTICLE{1983ApJ...275..713R,
       author = {{Rappaport}, S. and {Verbunt}, F. and {Joss}, P.~C.},
        title = "{A new technique for calculations of binary stellar evolution application to magnetic braking.}",
      journal = {\apj},
         year = 1983,
        month = dec,
       volume = {275},
        pages = {713-731},
          doi = {10.1086/161569},
       adsurl = {https://ui.adsabs.harvard.edu/abs/1983ApJ...275..713R}
}

@ARTICLE{1995ApJ...445..789P,
       author = {{Prialnik}, Dina and {Kovetz}, Attay},
        title = "{An Extended Grid of Multicycle Nova Evolution Models}",
      journal = {\apj},
         year = 1995,
        month = jun,
       volume = {445},
        pages = {789},
          doi = {10.1086/175741},
       adsurl = {https://ui.adsabs.harvard.edu/abs/1995ApJ...445..789P}
}

@ARTICLE{2005ApJ...623..398Y,
       author = {{Yaron}, O. and {Prialnik}, D. and {Shara}, M.~M. and {Kovetz}, A.},
        title = "{An Extended Grid of Nova Models. II. The Parameter Space of Nova Outbursts}",
      journal = {\apj},
         year = 2005,
        month = apr,
       volume = {623},
       number = {1},
        pages = {398-410},
          doi = {10.1086/428435},
archivePrefix = {arXiv},
       eprint = {astro-ph/0503143},
 primaryClass = {astro-ph},
       adsurl = {https://ui.adsabs.harvard.edu/abs/2005ApJ...623..398Y}
}

@ARTICLE{1995ApJ...439..330K,
       author = {{King}, A.~R. and {Kolb}, U.},
        title = "{Consequential Angular Momentum Loss and the Period Gap of Cataclysmic Variables}",
      journal = {\apj},
         year = 1995,
        month = jan,
       volume = {439},
        pages = {330},
          doi = {10.1086/175176},
       adsurl = {https://ui.adsabs.harvard.edu/abs/1995ApJ...439..330K}
}

@ARTICLE{2001ApJ...550..897H,
       author = {{Howell}, Steve B. and {Nelson}, Lorne A. and {Rappaport}, Saul},
        title = "{An Exploration of the Paradigm for the 2-3 Hour Period Gap in Cataclysmic Variables}",
      journal = {\apj},
         year = 2001,
        month = apr,
       volume = {550},
       number = {2},
        pages = {897-918},
          doi = {10.1086/319776},
archivePrefix = {arXiv},
       eprint = {astro-ph/0005435},
 primaryClass = {astro-ph},
       adsurl = {https://ui.adsabs.harvard.edu/abs/2001ApJ...550..897H}
}

@ARTICLE{1997A&A...327..620S,
       author = {{Soberman}, G.~E. and {Phinney}, E.~S. and {van den Heuvel}, E.~P.~J.},
        title = "{Stability criteria for mass transfer in binary stellar evolution.}",
      journal = {\aap},
         year = 1997,
        month = nov,
       volume = {327},
        pages = {620-635},
          doi = {10.48550/arXiv.astro-ph/9703016},
archivePrefix = {arXiv},
       eprint = {astro-ph/9703016},
 primaryClass = {astro-ph},
       adsurl = {https://ui.adsabs.harvard.edu/abs/1997A&A...327..620S}
}

@ARTICLE{2016MNRAS.455L..16S,
       author = {{Schreiber}, M.~R. and {Zorotovic}, M. and {Wijnen}, T.~P.~G.},
        title = "{Three in one go: consequential angular momentum loss can solve major problems of CV evolution}",
      journal = {\mnras},
         year = 2016,
        month = jan,
       volume = {455},
       number = {1},
        pages = {L16-L20},
          doi = {10.1093/mnrasl/slv144},
archivePrefix = {arXiv},
       eprint = {1510.04294},
 primaryClass = {astro-ph.SR},
       adsurl = {https://ui.adsabs.harvard.edu/abs/2016MNRAS.455L..16S}
}

@ARTICLE{2020AdSpR..66.1080Z,
       author = {{Zorotovic}, Monica and {Schreiber}, Matthias R.},
        title = "{Cataclysmic variable evolution and the white dwarf mass problem: A Review}",
      journal = {Advances in Space Research},
         year = 2020,
        month = sep,
       volume = {66},
       number = {5},
        pages = {1080-1089},
          doi = {10.1016/j.asr.2019.08.044},
archivePrefix = {arXiv},
       eprint = {1909.12323},
 primaryClass = {astro-ph.SR},
       adsurl = {https://ui.adsabs.harvard.edu/abs/2020AdSpR..66.1080Z}
}

@ARTICLE{2011A&A...536A..42Z,
       author = {{Zorotovic}, M. and {Schreiber}, M.~R. and {G{\"a}nsicke}, B.~T.},
        title = "{Post common envelope binaries from SDSS. XI. The white dwarf mass distributions of CVs and pre-CVs}",
      journal = {\aap},
         year = 2011,
        month = dec,
       volume = {536},
          eid = {A42},
        pages = {A42},
          doi = {10.1051/0004-6361/201116626},
archivePrefix = {arXiv},
       eprint = {1108.4600},
 primaryClass = {astro-ph.SR},
       adsurl = {https://ui.adsabs.harvard.edu/abs/2011A&A...536A..42Z}
}

@ARTICLE{2022MNRAS.510.6110P,
       author = {{Pala}, A.~F. and {G{\"a}nsicke}, B.~T. and {Belloni}, D. and {Parsons}, S.~G. and {Marsh}, T.~R. and {Schreiber}, M.~R. and {Breedt}, E. and {Knigge}, C. and {Sion}, E.~M. and {Szkody}, P. and {Townsley}, D. and {Bildsten}, L. and {Boyd}, D. and {Cook}, M.~J. and {De Martino}, D. and {Godon}, P. and {Kafka}, S. and {Kouprianov}, V. and {Long}, K.~S. and {Monard}, B. and {Myers}, G. and {Nelson}, P. and {Nogami}, D. and {Oksanen}, A. and {Pickard}, R. and {Poyner}, G. and {Reichart}, D.~E. and {Rodriguez Perez}, D. and {Shears}, J. and {Stubbings}, R. and {Toloza}, O.},
        title = "{Constraining the evolution of cataclysmic variables via the masses and accretion rates of their underlying white dwarfs}",
      journal = {\mnras},
         year = 2022,
        month = mar,
       volume = {510},
       number = {4},
        pages = {6110-6132},
          doi = {10.1093/mnras/stab3449},
archivePrefix = {arXiv},
       eprint = {2111.13706},
 primaryClass = {astro-ph.SR},
       adsurl = {https://ui.adsabs.harvard.edu/abs/2022MNRAS.510.6110P}
}

@ARTICLE{2017MNRAS.466L..63Z,
       author = {{Zorotovic}, M. and {Schreiber}, M.~R.},
        title = "{The origin of single low-mass WDs: another problem that consequential angular momentum loss in CVs might solve}",
      journal = {\mnras},
         year = 2017,
        month = mar,
       volume = {466},
       number = {1},
        pages = {L63-L67},
          doi = {10.1093/mnrasl/slw236},
archivePrefix = {arXiv},
       eprint = {1611.07309},
 primaryClass = {astro-ph.SR},
       adsurl = {https://ui.adsabs.harvard.edu/abs/2017MNRAS.466L..63Z}
}

@ARTICLE{2018MNRAS.478.5626B,
       author = {{Belloni}, Diogo and {Schreiber}, Matthias R. and {Zorotovic}, M{\'o}nica and {I{\l}kiewicz}, Krystian and {Hurley}, Jarrod R. and {Giersz}, Mirek and {Lagos}, Felipe},
        title = "{No cataclysmic variables missing: higher merger rate brings into agreement observed and predicted space densities}",
      journal = {\mnras},
         year = 2018,
        month = aug,
       volume = {478},
       number = {4},
        pages = {5626-5637},
          doi = {10.1093/mnras/sty1421},
archivePrefix = {arXiv},
       eprint = {1806.05687},
 primaryClass = {astro-ph.SR},
       adsurl = {https://ui.adsabs.harvard.edu/abs/2018MNRAS.478.5626B}
}

@ARTICLE{2022ApJ...938...31S,
       author = {{Shen}, Ken J. and {Quataert}, Eliot},
        title = "{Binary Interaction Dominates Mass Ejection in Classical Novae}",
      journal = {\apj},
         year = 2022,
        month = oct,
       volume = {938},
       number = {1},
          eid = {31},
        pages = {31},
          doi = {10.3847/1538-4357/ac9136},
archivePrefix = {arXiv},
       eprint = {2205.06283},
 primaryClass = {astro-ph.SR},
       adsurl = {https://ui.adsabs.harvard.edu/abs/2022ApJ...938...31S}
}

@ARTICLE{1986ApJ...305..251M,
       author = {{MacDonald}, J.},
        title = "{Postthermonuclear Runaway Angular Momentum Loss in Cataclysmic Binaries}",
      journal = {\apj},
         year = 1986,
        month = jun,
       volume = {305},
        pages = {251},
          doi = {10.1086/164245},
       adsurl = {https://ui.adsabs.harvard.edu/abs/1986ApJ...305..251M}
}

@ARTICLE{1986ApJ...311..163S,
       author = {{Shara}, Michael M. and {Livio}, Mario and {Moffat}, Anthony F.~J. and {Orio}, Marina},
        title = "{Do Novae Hibernate during Most of the Millennia between Eruptions? Links between Dwarf and Classical Novae, and Implications for the Space Densities and Evolution of Cataclysmic Binaries}",
      journal = {\apj},
         year = 1986,
        month = dec,
       volume = {311},
        pages = {163},
          doi = {10.1086/164762},
       adsurl = {https://ui.adsabs.harvard.edu/abs/1986ApJ...311..163S}
}

@ARTICLE{1998MNRAS.297..633S,
       author = {{Schenker}, K. and {Kolb}, U. and {Ritter}, H.},
        title = "{Properties of discontinuous and nova-amplified mass transfer in cataclysmic variables}",
      journal = {\mnras},
         year = 1998,
        month = jun,
       volume = {297},
       number = {2},
        pages = {633-647},
          doi = {10.1046/j.1365-8711.1998.01529.x},
archivePrefix = {arXiv},
       eprint = {astro-ph/9802138},
 primaryClass = {astro-ph},
       adsurl = {https://ui.adsabs.harvard.edu/abs/1998MNRAS.297..633S}
}

@ARTICLE{2019ApJ...870...22L,
       author = {{Liu}, Wei-Min and {Li}, Xiang-Dong},
        title = "{Can the Friction of the Nova Envelope Account for the Extra Angular Momentum Loss in Cataclysmic Variables?}",
      journal = {\apj},
         year = 2019,
        month = jan,
       volume = {870},
       number = {1},
          eid = {22},
        pages = {22},
          doi = {10.3847/1538-4357/aaf19f},
archivePrefix = {arXiv},
       eprint = {1811.08648},
 primaryClass = {astro-ph.SR},
       adsurl = {https://ui.adsabs.harvard.edu/abs/2019ApJ...870...22L}
}

@ARTICLE{2024ApJ...977...34T,
       author = {{Tang}, Wen-Shi and {Li}, Xiang-Dong and {Cui}, Zhe},
        title = "{Evolution of Cataclysmic Variables with Binary-driven Mass Loss during Nova Eruptions}",
      journal = {\apj},
         year = 2024,
        month = dec,
       volume = {977},
       number = {1},
          eid = {34},
        pages = {34},
          doi = {10.3847/1538-4357/ad8880},
archivePrefix = {arXiv},
       eprint = {2410.15254},
 primaryClass = {astro-ph.HE},
       adsurl = {https://ui.adsabs.harvard.edu/abs/2024ApJ...977...34T}
}

@ARTICLE{2026PhRvD.113l3023N,
       author = {{Nelson}, Nicholas and {Chang}, Philip},
        title = "{Moving-mesh simulations of minicommon envelope ejection in classical novae}",
      journal = {\prd},
         year = 2026,
        month = jun,
       volume = {113},
       number = {12},
          eid = {123023},
        pages = {123023},
          doi = {10.1103/jfbv-4lhm},
archivePrefix = {arXiv},
       eprint = {2604.24963},
 primaryClass = {astro-ph.HE},
       adsurl = {https://ui.adsabs.harvard.edu/abs/2026PhRvD.113l3023N}
}

@ARTICLE{1999ApJ...513..252O,
       author = {{Ostriker}, Eve C.},
        title = "{Dynamical Friction in a Gaseous Medium}",
      journal = {\apj},
         year = 1999,
        month = mar,
       volume = {513},
       number = {1},
        pages = {252-258},
          doi = {10.1086/306858},
archivePrefix = {arXiv},
       eprint = {astro-ph/9810324},
 primaryClass = {astro-ph},
       adsurl = {https://ui.adsabs.harvard.edu/abs/1999ApJ...513..252O}
}

@ARTICLE{1943ApJ....97..255C,
       author = {{Chandrasekhar}, S.},
        title = "{Dynamical Friction. I. General Considerations: the Coefficient of Dynamical Friction.}",
      journal = {\apj},
         year = 1943,
        month = mar,
       volume = {97},
        pages = {255},
          doi = {10.1086/144517},
       adsurl = {https://ui.adsabs.harvard.edu/abs/1943ApJ....97..255C}
}

@ARTICLE{2007ApJ...665..432K,
       author = {{Kim}, Hyosun and {Kim}, Woong-Tae},
        title = "{Dynamical Friction of a Circular-Orbit Perturber in a Gaseous Medium}",
      journal = {\apj},
         year = 2007,
        month = aug,
       volume = {665},
       number = {1},
        pages = {432-444},
          doi = {10.1086/519302},
archivePrefix = {arXiv},
       eprint = {0705.0084},
 primaryClass = {astro-ph},
       adsurl = {https://ui.adsabs.harvard.edu/abs/2007ApJ...665..432K}
}

@ARTICLE{2013ApJ...775...72B,
       author = {{Bernal}, Cristian G. and {S{\'a}nchez-Salcedo}, F.~J.},
        title = "{The Gravitational Drag Force on an Extended Object Moving in a Gas}",
      journal = {\apj},
         year = 2013,
        month = sep,
       volume = {775},
       number = {1},
          eid = {72},
        pages = {72},
          doi = {10.1088/0004-637X/775/1/72},
archivePrefix = {arXiv},
       eprint = {1308.4370},
 primaryClass = {astro-ph.CO},
       adsurl = {https://ui.adsabs.harvard.edu/abs/2013ApJ...775...72B}
}

@ARTICLE{1983ApJ...268..368E,
       author = {{Eggleton}, P.~P.},
        title = "{Aproximations to the radii of Roche lobes.}",
      journal = {\apj},
         year = 1983,
        month = may,
       volume = {268},
        pages = {368-369},
          doi = {10.1086/160960},
       adsurl = {https://ui.adsabs.harvard.edu/abs/1983ApJ...268..368E}
}

@ARTICLE{1999ApJ...522L..35S,
       author = {{S{\'a}nchez-Salcedo}, F.~J. and {Brandenburg}, A.},
        title = "{Deceleration by Dynamical Friction in a Gaseous Medium}",
      journal = {\apjl},
         year = 1999,
        month = sep,
       volume = {522},
       number = {1},
        pages = {L35-L38},
          doi = {10.1086/312215},
       adsurl = {https://ui.adsabs.harvard.edu/abs/1999ApJ...522L..35S}
}

@ARTICLE{2025MNRAS.541..396M,
       author = {{Morton}, B. and {Khochfar}, S. and {O{\~n}orbe}, J.},
        title = "{Gaseous dynamical friction: a numerical study of extended perturbers}",
      journal = {\mnras},
         year = 2025,
        month = jul,
       volume = {541},
       number = {1},
        pages = {396-408},
          doi = {10.1093/mnras/staf1016},
archivePrefix = {arXiv},
       eprint = {2103.15848},
 primaryClass = {astro-ph.GA},
       adsurl = {https://ui.adsabs.harvard.edu/abs/2025MNRAS.541..396M}
}

@ARTICLE{2024ApJ...966....7S,
       author = {{Suzuguchi}, Tomoya and {Sugimura}, Kazuyuki and {Hosokawa}, Takashi and {Matsumoto}, Tomoaki},
        title = "{Gas Dynamical Friction on Accreting Objects}",
      journal = {\apj},
         year = 2024,
        month = may,
       volume = {966},
       number = {1},
          eid = {7},
        pages = {7},
          doi = {10.3847/1538-4357/ad34af},
archivePrefix = {arXiv},
       eprint = {2401.13032},
 primaryClass = {astro-ph.GA},
       adsurl = {https://ui.adsabs.harvard.edu/abs/2024ApJ...966....7S}
}

@ARTICLE{2024ApJ...974..216O,
       author = {{O'Neill}, David and {D'Orazio}, Daniel J. and {Samsing}, Johan and {Pessah}, Martin E.},
        title = "{Gaseous Dynamical Friction on Elliptical Keplerian Orbits}",
      journal = {\apj},
         year = 2024,
        month = oct,
       volume = {974},
       number = {2},
          eid = {216},
        pages = {216},
          doi = {10.3847/1538-4357/ad7250},
archivePrefix = {arXiv},
       eprint = {2401.16166},
 primaryClass = {astro-ph.HE},
       adsurl = {https://ui.adsabs.harvard.edu/abs/2024ApJ...974..216O}
}

@ARTICLE{2026ApJ..1000..212O,
       author = {{O'Neill}, David and {D'Orazio}, Daniel J. and {Pessah}, Martin E.},
        title = "{Gaseous Dynamical Friction on Hyperbolic Scatterings}",
      journal = {\apj},
         year = 2026,
        month = apr,
       volume = {1000},
       number = {2},
          eid = {212},
        pages = {212},
          doi = {10.3847/1538-4357/ae48fc},
archivePrefix = {arXiv},
       eprint = {2505.20470},
 primaryClass = {astro-ph.GA},
       adsurl = {https://ui.adsabs.harvard.edu/abs/2026ApJ..1000..212O}
}

@ARTICLE{2013ApJ...777..136W,
       author = {{Wolf}, William M. and {Bildsten}, Lars and {Brooks}, Jared and {Paxton}, Bill},
        title = "{Hydrogen Burning on Accreting White Dwarfs: Stability, Recurrent Novae, and the Post-nova Supersoft Phase}",
      journal = {\apj},
         year = 2013,
        month = nov,
       volume = {777},
       number = {2},
          eid = {136},
        pages = {136},
          doi = {10.1088/0004-637X/777/2/136},
archivePrefix = {arXiv},
       eprint = {1309.3375},
 primaryClass = {astro-ph.SR},
       adsurl = {https://ui.adsabs.harvard.edu/abs/2013ApJ...777..136W}
}

@ARTICLE{2005A&A...439.1061S,
       author = {{Sala}, G. and {Hernanz}, M.},
        title = "{Models for the soft X-ray emission of post-outburst classical novae}",
      journal = {\aap},
         year = 2005,
        month = sep,
       volume = {439},
       number = {3},
        pages = {1061-1073},
          doi = {10.1051/0004-6361:20042622},
archivePrefix = {arXiv},
       eprint = {astro-ph/0504353},
 primaryClass = {astro-ph},
       adsurl = {https://ui.adsabs.harvard.edu/abs/2005A&A...439.1061S}
}

@ARTICLE{2021ARA&A..59..391C,
       author = {{Chomiuk}, Laura and {Metzger}, Brian D. and {Shen}, Ken J.},
        title = "{New Insights into Classical Novae}",
      journal = {\araa},
         year = 2021,
        month = sep,
       volume = {59},
        pages = {391-444},
          doi = {10.1146/annurev-astro-112420-114502},
archivePrefix = {arXiv},
       eprint = {2011.08751},
 primaryClass = {astro-ph.HE},
       adsurl = {https://ui.adsabs.harvard.edu/abs/2021ARA&A..59..391C}
}

@ARTICLE{2002MNRAS.329..897H,
       author = {{Hurley}, Jarrod R. and {Tout}, Christopher A. and {Pols}, Onno R.},
        title = "{Evolution of binary stars and the effect of tides on binary populations}",
      journal = {\mnras},
         year = 2002,
        month = feb,
       volume = {329},
       number = {4},
        pages = {897-928},
          doi = {10.1046/j.1365-8711.2002.05038.x},
archivePrefix = {arXiv},
       eprint = {astro-ph/0201220},
 primaryClass = {astro-ph},
       adsurl = {https://ui.adsabs.harvard.edu/abs/2002MNRAS.329..897H}
}

@PHDTHESIS{1989PhDT.........7H,
       author = {{Hjellming}, Michael Scott},
        title = "{Rapid Mass Transfer in Binary Systems.}",
       school = {University of Illinois, Urbana-Champaign},
         year = 1989,
        month = jan,
       adsurl = {https://ui.adsabs.harvard.edu/abs/1989PhDT.........7H}
}

@ARTICLE{2015ApJ...812...40G,
       author = {{Ge}, Hongwei and {Webbink}, Ronald F. and {Chen}, Xuefei and {Han}, Zhanwen},
        title = "{Adiabatic Mass Loss in Binary Stars. II. From Zero-age Main Sequence to the Base of the Giant Branch}",
      journal = {\apj},
         year = 2015,
        month = oct,
       volume = {812},
       number = {1},
          eid = {40},
        pages = {40},
          doi = {10.1088/0004-637X/812/1/40},
archivePrefix = {arXiv},
       eprint = {1507.04843},
 primaryClass = {astro-ph.SR},
       adsurl = {https://ui.adsabs.harvard.edu/abs/2015ApJ...812...40G}
}

@ARTICLE{2019MNRAS.486.5535M,
       author = {{McAllister}, M. and {Littlefair}, S.~P. and {Parsons}, S.~G. and {Dhillon}, V.~S. and {Marsh}, T.~R. and {G{\"a}nsicke}, B.~T. and {Breedt}, E. and {Copperwheat}, C. and {Green}, M.~J. and {Knigge}, C. and {Sahman}, D.~I. and {Dyer}, Martin J. and {Kerry}, P. and {Ashley}, R.~P. and {Irawati}, P. and {Rattanasoon}, S.},
        title = "{The evolutionary status of Cataclysmic Variables: eclipse modelling of 15 systems}",
      journal = {\mnras},
         year = 2019,
        month = jul,
       volume = {486},
       number = {4},
        pages = {5535-5551},
          doi = {10.1093/mnras/stz976},
archivePrefix = {arXiv},
       eprint = {1904.01888},
 primaryClass = {astro-ph.SR},
       adsurl = {https://ui.adsabs.harvard.edu/abs/2019MNRAS.486.5535M}
}

@ARTICLE{2024A&A...682A..33B,
       author = {{Belloni}, Diogo and {Schreiber}, Matthias R. and {Moe}, Maxwell and {El-Badry}, Kareem and {Shen}, Ken J.},
        title = "{Evidence for saturated and disrupted magnetic braking from samples of detached close binaries with M and K dwarfs}",
      journal = {\aap},
         year = 2024,
        month = feb,
       volume = {682},
          eid = {A33},
        pages = {A33},
          doi = {10.1051/0004-6361/202347931},
archivePrefix = {arXiv},
       eprint = {2311.04309},
 primaryClass = {astro-ph.SR},
       adsurl = {https://ui.adsabs.harvard.edu/abs/2024A&A...682A..33B}
}

@ARTICLE{2001MNRAS.322..231K,
       author = {{Kroupa}, Pavel},
        title = "{On the variation of the initial mass function}",
      journal = {\mnras},
         year = 2001,
        month = apr,
       volume = {322},
       number = {2},
        pages = {231-246},
          doi = {10.1046/j.1365-8711.2001.04022.x},
archivePrefix = {arXiv},
       eprint = {astro-ph/0009005},
 primaryClass = {astro-ph},
       adsurl = {https://ui.adsabs.harvard.edu/abs/2001MNRAS.322..231K}
}

@ARTICLE{2017ApJS..230...15M,
       author = {{Moe}, Maxwell and {Di Stefano}, Rosanne},
        title = "{Mind Your Ps and Qs: The Interrelation between Period (P) and Mass-ratio (Q) Distributions of Binary Stars}",
      journal = {\apjs},
         year = 2017,
        month = jun,
       volume = {230},
       number = {2},
          eid = {15},
        pages = {15},
          doi = {10.3847/1538-4365/aa6fb6},
archivePrefix = {arXiv},
       eprint = {1606.05347},
 primaryClass = {astro-ph.SR},
       adsurl = {https://ui.adsabs.harvard.edu/abs/2017ApJS..230...15M}
}

@ARTICLE{2016MNRAS.457.3867Z,
       author = {{Zorotovic}, M. and {Schreiber}, M.~R. and {Parsons}, S.~G. and {G{\"a}nsicke}, B.~T. and {Hardy}, A. and {Agurto-Gangas}, C. and {Nebot G{\'o}mez-Mor{\'a}n}, A. and {Rebassa-Mansergas}, A. and {Schwope}, A.~D.},
        title = "{Detached cataclysmic variables are crossing the orbital period gap}",
      journal = {\mnras},
         year = 2016,
        month = apr,
       volume = {457},
       number = {4},
        pages = {3867-3877},
          doi = {10.1093/mnras/stw246},
archivePrefix = {arXiv},
       eprint = {1601.07785},
 primaryClass = {astro-ph.SR},
       adsurl = {https://ui.adsabs.harvard.edu/abs/2016MNRAS.457.3867Z}
}

@ARTICLE{2010A&A...520A..86Z,
       author = {{Zorotovic}, M. and {Schreiber}, M.~R. and {G{\"a}nsicke}, B.~T. and {Nebot G{\'o}mez-Mor{\'a}n}, A.},
        title = "{Post-common-envelope binaries from SDSS. IX: Constraining the common-envelope efficiency}",
      journal = {\aap},
         year = 2010,
        month = sep,
       volume = {520},
          eid = {A86},
        pages = {A86},
          doi = {10.1051/0004-6361/200913658},
archivePrefix = {arXiv},
       eprint = {1006.1621},
 primaryClass = {astro-ph.SR},
       adsurl = {https://ui.adsabs.harvard.edu/abs/2010A&A...520A..86Z}
}

@ARTICLE{2023A&A...679L...8S,
       author = {{Schreiber}, Matthias R. and {Belloni}, Diogo and {van Roestel}, Jan},
        title = "{Period bouncers as detached magnetic cataclysmic variables}",
      journal = {\aap},
         year = 2023,
        month = nov,
       volume = {679},
          eid = {L8},
        pages = {L8},
          doi = {10.1051/0004-6361/202347766},
archivePrefix = {arXiv},
       eprint = {2310.17276},
 primaryClass = {astro-ph.SR},
       adsurl = {https://ui.adsabs.harvard.edu/abs/2023A&A...679L...8S}
}

@ARTICLE{2025A&A...698L..22S,
       author = {{Schreiber}, Matthias R. and {Belloni}, Diogo},
        title = "{The incidence of magnetic cataclysmic variables can be explained by the late appearance of white dwarf magnetic fields}",
      journal = {\aap},
         year = 2025,
        month = jun,
       volume = {698},
          eid = {L22},
        pages = {L22},
          doi = {10.1051/0004-6361/202554828},
archivePrefix = {arXiv},
       eprint = {2505.24153},
 primaryClass = {astro-ph.SR},
       adsurl = {https://ui.adsabs.harvard.edu/abs/2025A&A...698L..22S}
}

@ARTICLE{2020MNRAS.494.3799P,
       author = {{Pala}, A.~F. and {G{\"a}nsicke}, B.~T. and {Breedt}, E. and {Knigge}, C. and {Hermes}, J.~J. and {Gentile Fusillo}, N.~P. and {Hollands}, M.~A. and {Naylor}, T. and {Pelisoli}, I. and {Schreiber}, M.~R. and {Toonen}, S. and {Aungwerojwit}, A. and {Cukanovaite}, E. and {Dennihy}, E. and {Manser}, C.~J. and {Pretorius}, M.~L. and {Scaringi}, S. and {Toloza}, O.},
        title = "{A Volume-limited Sample of Cataclysmic Variables from Gaia DR2: Space Density and Population Properties}",
      journal = {\mnras},
         year = 2020,
        month = may,
       volume = {494},
       number = {3},
        pages = {3799-3827},
          doi = {10.1093/mnras/staa764},
archivePrefix = {arXiv},
       eprint = {1907.13152},
 primaryClass = {astro-ph.SR},
       adsurl = {https://ui.adsabs.harvard.edu/abs/2020MNRAS.494.3799P}
}

@ARTICLE{1972ApJ...175..417N,
       author = {{Nauenberg}, Michael},
        title = "{Analytic Approximations to the Mass-Radius Relation and Energy of Zero-Temperature Stars}",
      journal = {\apj},
         year = 1972,
        month = jul,
       volume = {175},
        pages = {417},
          doi = {10.1086/151568},
       adsurl = {https://ui.adsabs.harvard.edu/abs/1972ApJ...175..417N}
}

@ARTICLE{2020ApJ...905...62A,
       author = {{Aydi}, E. and {Chomiuk}, L. and {Izzo}, L. and {Harvey}, E.~J. and {Leahy-McGregor}, J. and {Strader}, J. and {Buckley}, D.~A.~H. and {Sokolovsky}, K.~V. and {Kawash}, A. and {Kochanek}, C.~S. and {Linford}, J.~D. and {Metzger}, B.~D. and {Mukai}, K. and {Orio}, M. and {Shappee}, B.~J. and {Shishkovsky}, L. and {Steinberg}, E. and {Swihart}, S.~J. and {Sokoloski}, J.~L. and {Walter}, F.~M. and {Woudt}, P.~A.},
        title = "{Early Spectral Evolution of Classical Novae: Consistent Evidence for Multiple Distinct Outflows}",
      journal = {\apj},
         year = 2020,
        month = dec,
       volume = {905},
       number = {1},
          eid = {62},
        pages = {62},
          doi = {10.3847/1538-4357/abc3bb},
archivePrefix = {arXiv},
       eprint = {2010.07481},
 primaryClass = {astro-ph.HE},
       adsurl = {https://ui.adsabs.harvard.edu/abs/2020ApJ...905...62A}
}

@ARTICLE{2011MNRAS.410..525M,
       author = {{Munari}, U. and {Ribeiro}, V.~A.~R.~M. and {Bode}, M.~F. and {Saguner}, T.},
        title = "{Properties, evolution and morpho-kinematical modelling of the very fast nova V2672 Oph (Nova Oph 2009), a clone of U Sco}",
      journal = {\mnras},
         year = 2011,
        month = jan,
       volume = {410},
       number = {1},
        pages = {525-534},
          doi = {10.1111/j.1365-2966.2010.17462.x},
archivePrefix = {arXiv},
       eprint = {1009.0334},
 primaryClass = {astro-ph.SR},
       adsurl = {https://ui.adsabs.harvard.edu/abs/2011MNRAS.410..525M}
}

@ARTICLE{2012A&A...543A..86N,
       author = {{Naito}, H. and {Mizoguchi}, S. and {Arai}, A. and {Tajitsu}, A. and {Narusawa}, S. and {Yamanaka}, M. and {Fujii}, M. and {Iijima}, T. and {Kinugasa}, K. and {Kurita}, M. and {Nagayama}, T. and {Yamaoka}, H. and {Sadakane}, K.},
        title = "{Five-year optical and near-infrared observations of the extremely slow nova V1280 Scorpii}",
      journal = {\aap},
         year = 2012,
        month = jul,
       volume = {543},
          eid = {A86},
        pages = {A86},
          doi = {10.1051/0004-6361/201118315},
archivePrefix = {arXiv},
       eprint = {1203.6725},
 primaryClass = {astro-ph.SR},
       adsurl = {https://ui.adsabs.harvard.edu/abs/2012A&A...543A..86N}
}

@ARTICLE{2016ApJ...833..149D,
       author = {{Darnley}, M.~J. and {Henze}, M. and {Bode}, M.~F. and {Hachisu}, I. and {Hernanz}, M. and {Hornoch}, K. and {Hounsell}, R. and {Kato}, M. and {Ness}, J.-U. and {Osborne}, J.~P. and {Page}, K.~L. and {Ribeiro}, V.~A.~R.~M. and {Rodr{\'\i}guez-Gil}, P. and {Shafter}, A.~W. and {Shara}, M.~M. and {Steele}, I.~A. and {Williams}, S.~C. and {Arai}, A. and {Arcavi}, I. and {Barsukova}, E.~A. and {Boumis}, P. and {Chen}, T. and {Fabrika}, S. and {Figueira}, J. and {Gao}, X. and {Gehrels}, N. and {Godon}, P. and {Goranskij}, V.~P. and {Harman}, D.~J. and {Hartmann}, D.~H. and {Hosseinzadeh}, G. and {Horst}, J. Chuck and {Itagaki}, K. and {Jos{\'e}}, J. and {Kabashima}, F. and {Kaur}, A. and {Kawai}, N. and {Kennea}, J.~A. and {Kiyota}, S. and {Ku{\v{c}}{\'a}kov{\'a}}, H. and {Lau}, K.~M. and {Maehara}, H. and {Naito}, H. and {Nakajima}, K. and {Nishiyama}, K. and {O'Brien}, T.~J. and {Quimby}, R. and {Sala}, G. and {Sano}, Y. and {Sion}, E.~M. and {Valeev}, A.~F. and {Watanabe}, F. and {Watanabe}, M. and {Williams}, B.~F. and {Xu}, Z.},
        title = "{M31N 2008-12a - The Remarkable Recurrent Nova in M31: Panchromatic Observations of the 2015 Eruption.}",
      journal = {\apj},
         year = 2016,
        month = dec,
       volume = {833},
       number = {2},
          eid = {149},
        pages = {149},
          doi = {10.3847/1538-4357/833/2/149},
archivePrefix = {arXiv},
       eprint = {1607.08082},
 primaryClass = {astro-ph.SR},
       adsurl = {https://ui.adsabs.harvard.edu/abs/2016ApJ...833..149D}
}

@ARTICLE{1994ApJ...437..802K,
       author = {{Kato}, Mariko and {Hachisu}, Izumi},
        title = "{Optically Thick Winds in Nova Outbursts}",
      journal = {\apj},
         year = 1994,
        month = dec,
       volume = {437},
        pages = {802},
          doi = {10.1086/175041},
       adsurl = {https://ui.adsabs.harvard.edu/abs/1994ApJ...437..802K}
}

@ARTICLE{2020A&A...634A..91B,
       author = {{Beuermann}, K. and {Burwitz}, V. and {Reinsch}, K. and {Schwope}, A. and {Thomas}, H.-C.},
        title = "{Neglected X-ray discovered polars. II. The peculiar eclipsing binary HY Eridani}",
      journal = {\aap},
         year = 2020,
        month = feb,
       volume = {634},
          eid = {A91},
        pages = {A91},
          doi = {10.1051/0004-6361/201936626},
archivePrefix = {arXiv},
       eprint = {2001.01967},
 primaryClass = {astro-ph.SR},
       adsurl = {https://ui.adsabs.harvard.edu/abs/2020A&A...634A..91B}
}

@ARTICLE{2019MNRAS.489.1023Y,
       author = {{Yu}, Zhifei and {Thorstensen}, J.~R. and {Rappaport}, S. and {Mann}, A. and {Jacobs}, T. and {Nelson}, L. and {G{\"a}nsicke}, B.~T. and {LaCourse}, D. and {Borkovits}, T. and {Aiken}, J. and {Steeghs}, D. and {Toloza}, O. and {Vanderburg}, A. and {Lin}, D.~N.~C.},
        title = "{A 9-h CV with one outburst in 4 yr of Kepler data}",
      journal = {\mnras},
         year = 2019,
        month = oct,
       volume = {489},
       number = {1},
        pages = {1023-1036},
          doi = {10.1093/mnras/stz2094},
archivePrefix = {arXiv},
       eprint = {1908.01914},
 primaryClass = {astro-ph.SR},
       adsurl = {https://ui.adsabs.harvard.edu/abs/2019MNRAS.489.1023Y}
}

@ARTICLE{2023MNRAS.525..785S,
       author = {{Schaefer}, Bradley E.},
        title = "{Orbital period changes for 14 novae and the critical failures of the predictions of standard theories, the Hibernation Model, and the Magnetic Braking Model}",
      journal = {\mnras},
         year = 2023,
        month = oct,
       volume = {525},
       number = {1},
        pages = {785-804},
          doi = {10.1093/mnras/stad2223},
archivePrefix = {arXiv},
       eprint = {2307.13804},
 primaryClass = {astro-ph.SR},
       adsurl = {https://ui.adsabs.harvard.edu/abs/2023MNRAS.525..785S}
}

@ARTICLE{2026A&A...710A.317D,
       author = {{Dodon}, Vladislav and {Li}, Xiang-Dong and {Xu}, Xiao-jie and {Galiullin}, Ilkham and {Sibgatullin}, Askar},
        title = "{Mechanisms for magnetic braking boost and disruption: The role of irradiation-driven winds and the convective turnover time spike in cataclysmic variables}",
      journal = {\aap},
         year = 2026,
        month = jun,
       volume = {710},
          eid = {A317},
        pages = {A317},
          doi = {10.1051/0004-6361/202660098},
archivePrefix = {arXiv},
       eprint = {2605.15897},
 primaryClass = {astro-ph.SR},
       adsurl = {https://ui.adsabs.harvard.edu/abs/2026A&A...710A.317D}
}

@ARTICLE{2024A&A...682L...7S,
       author = {{Schreiber}, Matthias R. and {Belloni}, Diogo and {Schwope}, Axel D.},
        title = "{The cataclysmic variable orbital period gap: More evident than ever}",
      journal = {\aap},
         year = 2024,
        month = feb,
       volume = {682},
          eid = {L7},
        pages = {L7},
          doi = {10.1051/0004-6361/202348807},
archivePrefix = {arXiv},
       eprint = {2402.02076},
 primaryClass = {astro-ph.SR},
       adsurl = {https://ui.adsabs.harvard.edu/abs/2024A&A...682L...7S}
}

@ARTICLE{2026arXiv260722836I,
       author = {{Inight}, K. and {G{\"a}nsicke}, B.~T. and {Aungwerojwit}, A. and {Izquierdo}, P. and {Manser}, C.~J. and {Myers}, A.~D. and {Swan}, A. and {Thorstensen}, J.~R. and {Aguilar}, J. and {Ahlen}, S. and {Bianchi}, D. and {Brooks}, D. and {Claybaugh}, T. and {de la Macorra}, A. and {Dey}, Biprateep and {Doel}, P. and {Font-Ribera}, A. and {Forero-Romero}, J.~E. and {Gontcho}, Satya Gontcho A and {Gutierrez}, G. and {Guy}, J. and {Joyce}, R. and {Juneau}, S. and {Koposov}, S.~E. and {Kremin}, A. and {Landriau}, M. and {Le Guillou}, L. and {Meisner}, A. and {Miquel}, R. and {Moustakas}, J. and {Najita}, J. and {P{\'e}rez-R{\`a}fols}, I. and {Percival}, W.~J. and {Prada}, F. and {Rossi}, G. and {Sanchez}, E. and {Schlegel}, D. and {Schubnell}, M. and {Seo}, H. and {Silber}, J. and {Tarl{\'e}}, G. and {Weaver}, B.~A. and {Zhou}, R. and {Zou}, H.},
        title = "{1000 cataclysmic variables identified from DESI spectroscopy}",
      journal = {arXiv e-prints},
         year = 2026,
        month = jul,
          eid = {arXiv:2607.22836},
        pages = {arXiv:2607.22836},
          doi = {10.48550/arXiv.2607.22836},
archivePrefix = {arXiv},
       eprint = {2607.22836},
 primaryClass = {astro-ph.SR},
       adsurl = {https://ui.adsabs.harvard.edu/abs/2026arXiv260722836I}
}

@ARTICLE{2026arXiv260728066S,
       author = {{Schwope}, A.~D. and {Brink}, J. and {Buckley}, D.~A.~H. and {C{\'u}neo}, V.~A. and {Friedrich}, S. and {Hern{\'a}ndez-D{\'\i}az}, S. and {Knauff}, K. and {Kurpas}, J. and {Lamer}, G. and {Potter}, S.~B. and {Pradeep}, K.~G. and {Schreiber}, M.~R. and {Stelzer}, B. and {Thorstensen}, J.~R. and {Tubin-Arenas}, D.},
        title = "{Cataclysmic variables from SRG/eROSITA: new systems from eRASS1}",
      journal = {arXiv e-prints},
         year = 2026,
        month = jul,
          eid = {arXiv:2607.28066},
        pages = {arXiv:2607.28066},
          doi = {10.48550/arXiv.2607.28066},
archivePrefix = {arXiv},
       eprint = {2607.28066},
 primaryClass = {astro-ph.SR},
       adsurl = {https://ui.adsabs.harvard.edu/abs/2026arXiv260728066S}
}

@ARTICLE{2026arXiv260727855H,
       author = {{Hern{\'a}ndez-D{\'\i}az}, S. and {Stelzer}, B. and {Schwope}, A. and {Mu{\~n}oz-Giraldo}, D. and {Schreiber}, M.~R. and {Brink}, J. and {Pradeep}, K.~G. and {G{\"a}nsicke}, B.~T. and {Eracleous}, M. and {Szekerczes}, K. and {Anderson}, S.~F. and {Brownstein}, J.~R.},
        title = "{Unveiling the period-bounce population of cataclysmic variables: Spectroscopic and time-domain follow-up of eROSITA-selected candidates}",
      journal = {arXiv e-prints},
         year = 2026,
        month = jul,
          eid = {arXiv:2607.27855},
        pages = {arXiv:2607.27855},
          doi = {10.48550/arXiv.2607.27855},
archivePrefix = {arXiv},
       eprint = {2607.27855},
 primaryClass = {astro-ph.SR},
       adsurl = {https://ui.adsabs.harvard.edu/abs/2026arXiv260727855H}
}

@ARTICLE{2026SSRv..222...32S,
       author = {{Scaringi}, Simone and {Knigge}, Christian and {de Martino}, Domitilla},
        title = "{Accreting White Dwarfs: An Unreview}",
      journal = {\ssr},
         year = 2026,
        month = mar,
       volume = {222},
       number = {3},
          eid = {32},
        pages = {32},
          doi = {10.1007/s11214-026-01290-x},
archivePrefix = {arXiv},
       eprint = {2603.10150},
 primaryClass = {astro-ph.HE},
       adsurl = {https://ui.adsabs.harvard.edu/abs/2026SSRv..222...32S}
}

@ARTICLE{1993A&A...271..149K,
       author = {{Kolb}, U.},
        title = "{A model for the intrinsic population of cataclysmic variables}",
      journal = {\aap},
         year = 1993,
        month = apr,
       volume = {271},
        pages = {149},
       adsurl = {https://ui.adsabs.harvard.edu/abs/1993A&A...271..149K}
}

@ARTICLE{2009MNRAS.397.2170G,
       author = {{G{\"a}nsicke}, B.~T. and {Dillon}, M. and {Southworth}, J. and {Thorstensen}, J.~R. and {Rodr{\'\i}guez-Gil}, P. and {Aungwerojwit}, A. and {Marsh}, T.~R. and {Szkody}, P. and {Barros}, S.~C.~C. and {Casares}, J. and {de Martino}, D. and {Groot}, P.~J. and {Hakala}, P. and {Kolb}, U. and {Littlefair}, S.~P. and {Mart{\'\i}nez-Pais}, I.~G. and {Nelemans}, G. and {Schreiber}, M.~R.},
        title = "{SDSS unveils a population of intrinsically faint cataclysmic variables at the minimum orbital period}",
      journal = {\mnras},
         year = 2009,
        month = aug,
       volume = {397},
       number = {4},
        pages = {2170-2188},
          doi = {10.1111/j.1365-2966.2009.15126.x},
archivePrefix = {arXiv},
       eprint = {0905.3476},
 primaryClass = {astro-ph.SR},
       adsurl = {https://ui.adsabs.harvard.edu/abs/2009MNRAS.397.2170G}
}

@ARTICLE{2012MNRAS.419.1442P,
       author = {{Pretorius}, Magaretha L. and {Knigge}, Christian},
        title = "{The space density and X-ray luminosity function of non-magnetic cataclysmic variables}",
      journal = {\mnras},
         year = 2012,
        month = jan,
       volume = {419},
       number = {2},
        pages = {1442-1454},
          doi = {10.1111/j.1365-2966.2011.19801.x},
archivePrefix = {arXiv},
       eprint = {1109.3162},
 primaryClass = {astro-ph.SR},
       adsurl = {https://ui.adsabs.harvard.edu/abs/2012MNRAS.419.1442P}
}

@ARTICLE{2015ApJ...805L...6S,
       author = {{Shen}, Ken J.},
        title = "{Every Interacting Double White Dwarf Binary May Merge}",
      journal = {\apjl},
         year = 2015,
        month = may,
       volume = {805},
       number = {1},
          eid = {L6},
        pages = {L6},
          doi = {10.1088/2041-8205/805/1/L6},
archivePrefix = {arXiv},
       eprint = {1502.05052},
 primaryClass = {astro-ph.SR},
       adsurl = {https://ui.adsabs.harvard.edu/abs/2015ApJ...805L...6S}
}

@ARTICLE{2021ApJ...914....5S,
       author = {{Sparks}, Warren M. and {Sion}, Edward M.},
        title = "{Nova-produced Common Envelope: Source of the Nonsolar Abundances and an Additional Frictional Angular Momentum Loss in Cataclysmic Variables}",
      journal = {\apj},
         year = 2021,
        month = jun,
       volume = {914},
       number = {1},
          eid = {5},
        pages = {5},
          doi = {10.3847/1538-4357/abf2bc},
archivePrefix = {arXiv},
       eprint = {2103.14149},
 primaryClass = {astro-ph.SR},
       adsurl = {https://ui.adsabs.harvard.edu/abs/2021ApJ...914....5S}
}

@ARTICLE{1991A&A...246...84L,
       author = {{Livio}, M. and {Govarie}, A. and {Ritter}, H.},
        title = "{Changes in the binary system parameters resulting from nova explosions}",
      journal = {\aap},
         year = 1991,
        month = jun,
       volume = {246},
       number = {1},
        pages = {84-90},
       adsurl = {https://ui.adsabs.harvard.edu/abs/1991A&A...246...84L}
}

@ARTICLE{2025A&A...696A..92B,
       author = {{Barraza-Jorquera}, Joaqu{\'\i}n A. and {Schreiber}, Matthias R. and {Belloni}, Diogo},
        title = "{Further evidence of saturated, boosted, and disrupted magnetic braking from evolutionary tracks of cataclysmic variables}",
      journal = {\aap},
         year = 2025,
        month = apr,
       volume = {696},
          eid = {A92},
        pages = {A92},
          doi = {10.1051/0004-6361/202553757},
archivePrefix = {arXiv},
       eprint = {2503.18884},
 primaryClass = {astro-ph.SR},
       adsurl = {https://ui.adsabs.harvard.edu/abs/2025A&A...696A..92B}
}

@ARTICLE{2026A&A...708L..11B,
       author = {{Barraza-Jorquera}, Joaqu{\'\i}n A. and {Schreiber}, Matthias R. and {Littlefair}, Stuart and {Belloni}, Diogo and {Schwope}, Axel D.},
        title = "{A modest change in magnetic braking at the fully convective boundary explains the evolution of cataclysmic variables}",
      journal = {\aap},
         year = 2026,
        month = apr,
       volume = {708},
          eid = {L11},
        pages = {L11},
          doi = {10.1051/0004-6361/202659326},
archivePrefix = {arXiv},
       eprint = {2603.14560},
 primaryClass = {astro-ph.SR},
       adsurl = {https://ui.adsabs.harvard.edu/abs/2026A&A...708L..11B}
}

@ARTICLE{2026PASA...43....2B,
       author = {{Bhattacharyya}, Soumik and {Chamandy}, Luke and {Blackman}, Eric G. and {Frank}, Adam and {Liu}, Baowei},
        title = "{Understanding the drag torque in common envelope evolution}",
      journal = {\pasa},
         year = 2026,
        month = jan,
       volume = {43},
          eid = {e002},
        pages = {e002},
          doi = {10.1017/pasa.2025.10133},
archivePrefix = {arXiv},
       eprint = {2506.19547},
 primaryClass = {astro-ph.SR},
       adsurl = {https://ui.adsabs.harvard.edu/abs/2026PASA...43....2B}
}

@ARTICLE{2008ApJ...679L..33K,
       author = {{Kim}, Hyosun and {Kim}, Woong-Tae and {S{\'a}nchez-Salcedo}, F.~J.},
        title = "{Dynamical Friction of Double Perturbers in a Gaseous Medium}",
      journal = {\apjl},
         year = 2008,
        month = may,
       volume = {679},
       number = {1},
        pages = {L33},
          doi = {10.1086/589149},
archivePrefix = {arXiv},
       eprint = {0804.2010},
 primaryClass = {astro-ph},
       adsurl = {https://ui.adsabs.harvard.edu/abs/2008ApJ...679L..33K}
}

@ARTICLE{2026NatAs..10..271A,
       author = {{Aydi}, Elias and {Monnier}, John D. and {M{\'e}rand}, Antoine and {Schaefer}, Gail H. and {Chomiuk}, Laura and {Otulakowska-Hypka}, Magdalena and {Fan}, Jhih-Ling and {Li}, Kwan Lok and {Sokolovsky}, Kirill V. and {Salinas}, Ricardo and {Tucker}, Michael and {Shappee}, Benjamin and {Rudy}, Richard and {Page}, Kim L. and {Kuin}, N. Paul M. and {Buckley}, David A.~H. and {Craig}, Peter and {Izzo}, Luca and {Linford}, Justin and {Metzger}, Brian D. and {Mukai}, Koji and {Orio}, Marina and {Shen}, Ken J. and {Strader}, Jay and {Sokoloski}, Jennifer L. and {Williams}, Robert E. and {Williams}, Montana N. and {Habtie}, Gesesew R. and {Kraus}, Stefan and {Anugu}, Narsireddy and {Bouquin}, Jean-Baptiste Le and {Chhabra}, Sorabh and {Codron}, Isabelle and {Gardner}, Tyler and {Gutierrez}, Mayra and {Ibrahim}, Noura and {Lanthermann}, Cyprien and {Setterholm}, Benjamin R. and {Ashall}, Christopher and {Hinkle}, Jason T. and {Jaeger}, Thomas de and {Payne}, Anna V.},
        title = "{Multiple outflows and delayed ejections revealed by early imaging of novae}",
      journal = {Nature Astronomy},
         year = 2026,
        month = feb,
       volume = {10},
        pages = {271-280},
          doi = {10.1038/s41550-025-02725-1},
archivePrefix = {arXiv},
       eprint = {2512.05220},
 primaryClass = {astro-ph.HE},
       adsurl = {https://ui.adsabs.harvard.edu/abs/2026NatAs..10..271A}
}

@ARTICLE{2011A&A...536A..97F,
       author = {{Friedjung}, M.},
        title = "{Do novae have optically thick winds during outburst with large deviations from spherical symmetry?}",
      journal = {\aap},
         year = 2011,
        month = dec,
       volume = {536},
          eid = {A97},
        pages = {A97},
          doi = {10.1051/0004-6361/201016397},
archivePrefix = {arXiv},
       eprint = {1201.4561},
 primaryClass = {astro-ph.SR},
       adsurl = {https://ui.adsabs.harvard.edu/abs/2011A&A...536A..97F}
}

@ARTICLE{2021ApJ...923..100M,
       author = {{Metzger}, Brian D. and {Zenati}, Yossef and {Chomiuk}, Laura and {Shen}, Ken J. and {Strader}, Jay},
        title = "{Transients from the Cataclysmic Deaths of Cataclysmic Variables}",
      journal = {\apj},
         year = 2021,
        month = dec,
       volume = {923},
       number = {1},
          eid = {100},
        pages = {100},
          doi = {10.3847/1538-4357/ac2a39},
archivePrefix = {arXiv},
       eprint = {2108.04305},
 primaryClass = {astro-ph.SR},
       adsurl = {https://ui.adsabs.harvard.edu/abs/2021ApJ...923..100M}
}

@ARTICLE{2018MNRAS.474..427Z,
       author = {{Zhang}, Xianfei and {Hall}, Philip D. and {Jeffery}, C. Simon and {Bi}, Shaolan},
        title = "{Evolution models of helium white dwarf-main-sequence star merger remnants: the mass distribution of single low-mass white dwarfs}",
      journal = {\mnras},
         year = 2018,
        month = feb,
       volume = {474},
       number = {1},
        pages = {427-432},
          doi = {10.1093/mnras/stx2747},
archivePrefix = {arXiv},
       eprint = {1711.03285},
 primaryClass = {astro-ph.SR},
       adsurl = {https://ui.adsabs.harvard.edu/abs/2018MNRAS.474..427Z}
}

@ARTICLE{2012ApJ...745...43N,
       author = {{Ness}, J.-U. and {Schaefer}, B.~E. and {Dobrotka}, A. and {Sadowski}, A. and {Drake}, J.~J. and {Barnard}, R. and {Talavera}, A. and {Gonzalez-Riestra}, R. and {Page}, K.~L. and {Hernanz}, M. and {Sala}, G. and {Starrfield}, S.},
        title = "{From X-Ray Dips to Eclipse: Witnessing Disk Reformation in the Recurrent Nova U Sco}",
      journal = {\apj},
         year = 2012,
        month = jan,
       volume = {745},
       number = {1},
          eid = {43},
        pages = {43},
          doi = {10.1088/0004-637X/745/1/43},
archivePrefix = {arXiv},
       eprint = {1105.2717},
 primaryClass = {astro-ph.GA},
       adsurl = {https://ui.adsabs.harvard.edu/abs/2012ApJ...745...43N}
}

@ARTICLE{2018A&A...613A...8F,
       author = {{Figueira}, Joana and {Jos{\'e}}, Jordi and {Garc{\'\i}a-Berro}, Enrique and {Campbell}, Simon W. and {Garc{\'\i}a-Senz}, Domingo and {Mohamed}, Shazrene},
        title = "{Three-dimensional simulations of the interaction between the nova ejecta, accretion disk, and companion star}",
      journal = {\aap},
         year = 2018,
        month = may,
       volume = {613},
          eid = {A8},
        pages = {A8},
          doi = {10.1051/0004-6361/201731545},
archivePrefix = {arXiv},
       eprint = {1712.08402},
 primaryClass = {astro-ph.SR},
       adsurl = {https://ui.adsabs.harvard.edu/abs/2018A&A...613A...8F}
}

@ARTICLE{2026arXiv260716422G,
       author = {{Gagnier}, Damien},
        title = "{Dynamical friction in stratified stellar envelopes}",
      journal = {arXiv e-prints},
         year = 2026,
        month = jul,
          eid = {arXiv:2607.16422},
        pages = {arXiv:2607.16422},
          doi = {10.48550/arXiv.2607.16422},
archivePrefix = {arXiv},
       eprint = {2607.16422},
 primaryClass = {astro-ph.SR},
       adsurl = {https://ui.adsabs.harvard.edu/abs/2026arXiv260716422G}
}

@ARTICLE{2011ApJS..192....3P,
       author = {{Paxton}, Bill and {Bildsten}, Lars and {Dotter}, Aaron and {Herwig}, Falk and {Lesaffre}, Pierre and {Timmes}, Frank},
        title = "{Modules for Experiments in Stellar Astrophysics (MESA)}",
      journal = {\apjs},
         year = 2011,
        month = jan,
       volume = {192},
       number = {1},
          eid = {3},
        pages = {3},
          doi = {10.1088/0067-0049/192/1/3},
archivePrefix = {arXiv},
       eprint = {1009.1622},
 primaryClass = {astro-ph.SR},
       adsurl = {https://ui.adsabs.harvard.edu/abs/2011ApJS..192....3P}
}

@ARTICLE{2013ApJS..208....4P,
       author = {{Paxton}, Bill and {Cantiello}, Matteo and {Arras}, Phil and {Bildsten}, Lars and {Brown}, Edward F. and {Dotter}, Aaron and {Mankovich}, Christopher and {Montgomery}, M.~H. and {Stello}, Dennis and {Timmes}, F.~X. and et al.},
        title = "{Modules for Experiments in Stellar Astrophysics (MESA): Planets, Oscillations, Rotation, and Massive Stars}",
      journal = {\apjs},
         year = 2013,
        month = sep,
       volume = {208},
       number = {1},
          eid = {4},
        pages = {4},
          doi = {10.1088/0067-0049/208/1/4},
archivePrefix = {arXiv},
       eprint = {1301.0319},
 primaryClass = {astro-ph.SR},
       adsurl = {https://ui.adsabs.harvard.edu/abs/2013ApJS..208....4P}
}

@ARTICLE{2015ApJS..220...15P,
       author = {{Paxton}, Bill and {Marchant}, Pablo and {Schwab}, Josiah and {Bauer}, Evan B. and {Bildsten}, Lars and {Cantiello}, Matteo and {Dessart}, Luc and {Farmer}, R. and {Hu}, H. and {Langer}, N. and et al.},
        title = "{Modules for Experiments in Stellar Astrophysics (MESA): Binaries, Pulsations, and Explosions}",
      journal = {\apjs},
         year = 2015,
        month = sep,
       volume = {220},
       number = {1},
          eid = {15},
        pages = {15},
          doi = {10.1088/0067-0049/220/1/15},
archivePrefix = {arXiv},
       eprint = {1506.03146},
 primaryClass = {astro-ph.SR},
       adsurl = {https://ui.adsabs.harvard.edu/abs/2015ApJS..220...15P}
}

@ARTICLE{2018ApJS..234...34P,
       author = {{Paxton}, Bill and {Schwab}, Josiah and {Bauer}, Evan B. and {Bildsten}, Lars and {Blinnikov}, Sergei and {Duffell}, Paul and {Farmer}, R. and {Goldberg}, Jared A. and {Marchant}, Pablo and {Sorokina}, Elena and et al.},
        title = "{Modules for Experiments in Stellar Astrophysics (MESA): Convective Boundaries, Element Diffusion, and Massive Star Explosions}",
      journal = {\apjs},
         year = 2018,
        month = feb,
       volume = {234},
       number = {2},
          eid = {34},
        pages = {34},
          doi = {10.3847/1538-4365/aaa5a8},
archivePrefix = {arXiv},
       eprint = {1710.08424},
 primaryClass = {astro-ph.SR},
       adsurl = {https://ui.adsabs.harvard.edu/abs/2018ApJS..234...34P}
}

@ARTICLE{2019ApJS..243...10P,
       author = {{Paxton}, Bill and {Smolec}, R. and {Schwab}, Josiah and {Gautschy}, A. and {Bildsten}, Lars and {Cantiello}, Matteo and {Dotter}, Aaron and {Farmer}, R. and {Goldberg}, Jared A. and {Jermyn}, Adam S. and et al.},
        title = "{Modules for Experiments in Stellar Astrophysics (MESA): Pulsating Variable Stars, Rotation, Convective Boundaries, and Energy Conservation}",
      journal = {\apjs},
         year = 2019,
        month = jul,
       volume = {243},
       number = {1},
          eid = {10},
        pages = {10},
          doi = {10.3847/1538-4365/ab2241},
archivePrefix = {arXiv},
       eprint = {1903.01426},
 primaryClass = {astro-ph.SR},
       adsurl = {https://ui.adsabs.harvard.edu/abs/2019ApJS..243...10P}
}

@ARTICLE{2023ApJS..265...15J,
       author = {{Jermyn}, Adam S. and {Bauer}, Evan B. and {Schwab}, Josiah and {Farmer}, R. and {Ball}, Warrick H. and {Bellinger}, Earl P. and {Dotter}, Aaron and {Joyce}, Meridith and {Marchant}, Pablo and {Mombarg}, Joey S.~G. and et al.},
        title = "{Modules for Experiments in Stellar Astrophysics (MESA): Time-dependent Convection, Energy Conservation, Automatic Differentiation, and Infrastructure}",
      journal = {\apjs},
         year = 2023,
        month = mar,
       volume = {265},
       number = {1},
          eid = {15},
        pages = {15},
          doi = {10.3847/1538-4365/acae8d},
archivePrefix = {arXiv},
       eprint = {2208.03651},
 primaryClass = {astro-ph.SR},
       adsurl = {https://ui.adsabs.harvard.edu/abs/2023ApJS..265...15J}
}

@ARTICLE{1988A&A...202...93R,
       author = {{Ritter}, H.},
        title = "{Turning on and off mass transfer in cataclysmic binaries.}",
      journal = {\aap},
         year = 1988,
        month = aug,
       volume = {202},
        pages = {93-100},
       adsurl = {https://ui.adsabs.harvard.edu/abs/1988A&A...202...93R}
}

@ARTICLE{2020MNRAS.491.5717B,
       author = {{Belloni}, Diogo and {Schreiber}, Matthias R. and {Pala}, Anna F. and {G{\"a}nsicke}, Boris T. and {Zorotovic}, M{\'o}nica and {Rodrigues}, Claudia V.},
        title = "{Evidence for reduced magnetic braking in polars from binary population models}",
      journal = {\mnras},
         year = 2020,
        month = feb,
       volume = {491},
       number = {4},
        pages = {5717-5731},
          doi = {10.1093/mnras/stz3413},
archivePrefix = {arXiv},
       eprint = {1910.06333},
 primaryClass = {astro-ph.SR},
       adsurl = {https://scixplorer.org/abs/2020MNRAS.491.5717B}
}

@ARTICLE{2022MNRAS.511.4060K,
       author = {{Kamlah}, A.~W.~H. and {Leveque}, A. and {Spurzem}, R. and {Arca Sedda}, M. and {Askar}, A. and {Banerjee}, S. and {Berczik}, P. and {Giersz}, M. and {Hurley}, J. and {Belloni}, D. and et al.},
        title = "{Preparing the next gravitational million-body simulations: evolution of single and binary stars in NBODY6++GPU, MOCCA, and MCLUSTER}",
      journal = {\mnras},
         year = 2022,
        month = apr,
       volume = {511},
       number = {3},
        pages = {4060-4089},
          doi = {10.1093/mnras/stab3748},
archivePrefix = {arXiv},
       eprint = {2105.08067},
 primaryClass = {astro-ph.GA},
       adsurl = {https://scixplorer.org/abs/2022MNRAS.511.4060K}
}
\bibliographystyle{aasjournalv7}

\end{document}